\documentclass[12pt]{article}
\pdfoutput=1

\usepackage[pdftex]{graphicx}
\usepackage{amssymb}
\usepackage{amsmath}
\usepackage{cite}
\usepackage{multirow}

\usepackage{longtable}
\usepackage{booktabs}
\usepackage{array}

\usepackage{wrapfig}
\usepackage{color}

\renewcommand{\arraystretch}{1.2}

\newcommand{\av}[1]{\langle #1 \rangle}

\newcommand{\nn}{\nonumber}

\definecolor{DRed}{rgb}{0.8,0,0.1}
\definecolor{DBlue}{rgb}{0,0,0.8}
\definecolor{DGreen}{rgb}{0,0.5,0.3}
\definecolor{DPurple}{rgb}{1.0,0,0.8}

\begin{document}

\begin{flushright}
{\small
 ICCUB-17-003\\ 
 UAB-FT-990 \\ 
LPT-ORSAY/16-89\\

}
\end{flushright}
$\ $
\vspace{-2mm}
\begin{center}
\Large \bf  
Hadronic uncertainties in $B\to K^*\mu^+\mu^-$:\\ a state-of-the-art analysis
\end{center}

\vspace{1mm}
\begin{center}
{\sf Bernat Capdevila$^a$, S\'ebastien Descotes-Genon$^b$, Lars Hofer$^c$, Joaquim Matias$^a$}\\[5mm]
{\em \small
$^a$Universitat Aut\`onoma de Barcelona, 08193 Bellaterra, Barcelona,\\
Institut de Fisica d'Altes Energies (IFAE), The Barcelona Institute of Science and Technology, Campus UAB, 08193 Bellaterra (Barcelona) Spain
}\\[5mm]
{\em\small
$^b$ Laboratoire de Physique Th\'eorique (UMR8627),\\
CNRS, Univ. Paris-Sud, Universit\'e Paris-Saclay, 91405 Orsay, France}
\\[5mm]
{\em \small
$^c$ Department de F\'isica Qu\`antica i Astrof\'isica, Institut de Ci\`encies del Cosmos,\\
Universitat de Barcelona, 08028 Barcelona, Spain
}
\end{center}

\vspace{1mm}
\begin{abstract}
\bigskip

\noindent

In the absence of direct evidence for New Physics at present LHC energies,
the focus is set on the anomalies and discrepancies recently observed
in rare $b \to s\ell\ell$ transitions which can be interpreted as indirect hints. Global fits have shown that an economical New Physics solution can simultaneously alleviate the tensions in the various channels and can lead to a significant improvement in the description of the data. Alternative explanations within the Standard Model for 
part of the observed anomalies have been proposed  in terms of (unexpectedly large) hadronic effects  at low dilepton invariant mass and attributing tensions in protected observables to statistical fluctuations or experimental errors.
We review the treatment of hadronic uncertainties in this kinematic regime for one of the most important channels, $B \to K^*\mu^+\mu^-$, in a pedagogical way.
We provide detailed arguments showing that factorisable power corrections cannot account for the observed anomalies and that an explanation through
long-distance charm contributions is disfavoured.  
Some optimized observables at very low dilepton invariant mass are shown to be protected against contributions from the semileptonic coefficient $C_9$ (including any associated long-distance charm effects),
enhancing their sensitivity to New Physics contributions to other Wilson coefficients.  
Finally, we discuss how the recent measurement of $Q_5$ by Belle (and in the future by LHCb and Belle-II) may provide a robust cross-check of our arguments. 

\end{abstract}

\setcounter{tocdepth}{2}

\newpage
\section{Introduction}
\label{sec:intro}
For many years the Standard Model (SM) has been probed and systematically confirmed in collider experiments, 
with tensions showing up only temporarily and in isolated channels.
However, in recent years a consistent picture of tensions has emerged in interrelated channels in the flavour sector.
In the 1 fb$^{-1}$ data set~\cite{Aaij:2013qta}, evaluated in 2013, LHCb detected a sizeable 3.7 $\sigma$ deviation in 
one bin of the angular observable $P_5^\prime$\cite{DescotesGenon:2012zf} in the decay $B\to K^*\mu^+\mu^-$ 
(the so-called $P_5^\prime$ anomaly \cite{Descotes-Genon:2013wba}).
The fact that this anomaly was accompanied by a 2.9 $\sigma$ tension in the second 
bin of 
another angular observable called $P_2$ (related to the forward-backward asymmetry)~\footnote{Unfortunately, this tension observed in 2013 cannot be seen in the 2015 data due to
a combination of circumstances, namely a change in the binning of LHCb data
together with a measured value of $F_L$ compatible with $F_L=1$ within errors in the third bin. A more precise measurement of $F_L$, accessible 
with more statistics, should help recovering this important piece of information in the future.}  pointed, for the first time, to a coherent pattern of deviations
\cite{Descotes-Genon:2013wba}. In 2015, using the 3 fb$^{-1}$ data set~\cite{Aaij:2015oid}, LHCb provided more accurate results for these angular observables, once again with a discrepancy betwen the measurement of $P_5^\prime$ and the theory prediction within the SM. The same experiment also uncovered new deviations (larger than $2\sigma$) in the  $B_s \to \phi \mu^+\mu^-$ branching ratio (at low and large $\phi$ recoil)~\cite{Aaij:2013aln,Aaij:2015esa}. A few months ago, the Belle experiment performed an independent measurement of $P_5^\prime$~\cite{Abdesselam:2016llu}: the value, compatible with the LHCb measurements, agrees again poorly with the theory expectations in the SM.

Another interesting tension  was observed in the ratio $R_K={\cal B}_{B \to K \mu^+\mu^-}$ $/{\cal B}_{B \to K e^+e^-}$ indicating that this deviation would affect predominantly
$b\to s\mu^+\mu^-$ compared to $b\to s e^+e^-$~\cite{Aaij:2014ora}, and thus violate lepton-flavour universality. This difference among lepton modes was also supported by the fact that no deviation was observed in $B\to K^* e^+e^-$ data at very large $K^*$ recoil~\cite{Aaij:2015dea}.
Very recently, Belle has presented a separate measurement~\cite{Wehle:2016yoi} of $P_5^\prime$ in the muon and electron channels,
and hence of the observable $Q_5=P_{5}^{\mu\prime}-P_{5}^{e\prime}$ proposed in Ref.~\cite{Capdevila:2016ivx}. 
While the muon channel exhibits a 2.6 $\sigma$ deviation with respect to the SM prediction~\cite{Descotes-Genon:2015uva} and in good agreement
with the LHCb measurement, the electron channel 
agrees with the SM expectation at 1.3 $\sigma$. Though it is not yet statistically significant,
the result could point to a violation of lepton-flavour universality in $P_5^\prime$ in compliance with the one measured in $R_K$. If this result is confirmed by LHCb with higher statistics and also other tensions in new experimental measurements of lepton flavour universality ratios, like the promising $R_{K^*}={\cal B}_{B \to K^* \mu^+\mu^-}$ $/{\cal B}_{B \to K^* e^+e^-}$~\cite{Hiller:2003js,Capdevila:2016ivx,Descotes-Genon:2015uva}, are detected, this would hamper any attempt to explain the $P_5^\prime$ anomaly in terms of non-perturbative QCD effects. 

It is striking that all the above-mentioned deviations can be alleviated simultaneously by a common mechanism, namely by
a New Physics (NP) contribution to the short-distance coefficient of the semi-leptonic operator ${\cal O}_9^\mu$, i.e. to the  vector component of the $b\to s\mu\mu$ transition 
in  the effective Hamiltonian describing these transitions at the $b$-quark scale.
Global analyses of  $b \to s \ell \ell$ decays performed by independent groups \cite{Descotes-Genon:2015uva,Altmannshofer:2014rta,Altmannshofer:2015sma,Hurth:2016fbr} following different approaches (improved QCD-Factorisation or full form factors), using different form factor input 
(from Ref.~\cite{Khodjamirian:2010vf} or \cite{Straub:2015ica}) and different observables (optimized $P^{(\prime)}_i$ or form factor dependent $S_i$) have established that a negative New Physics (NP) contribution to the Wilson coefficient $C_9^\mu$ of $\sim-$25\% with respect to its SM value is favoured with large significance (between 4-5 $\sigma$ depending on the hypothesis on the Wilson coefficients receiving NP contributions).

However, a controversy arose concerning the interpretation of the observed deviations in the semi-leptonic $B_{(s)}$ decays since the predictions are plagued by perturbative and non-perturbative QCD effects and some of the non-perturbative effects may mimic a NP signal. It was argued that unexpectedly large effects could be caused by resonance tails leaking into the $q^2<8$ GeV$^2$ region. Very recently, LHCb measured the relative phases
of the $J/\psi$ and $\psi(2S)$ with the short-distance contribution to $B\to K\mu^+\mu^-$ and reported small interference effects in dimuon mass regions far from the pole masses of the resonances~\cite{Aaij:2016cbx}. The obtained fit is coherent with the global analyses~\cite{Descotes-Genon:2015uva,Altmannshofer:2014rta,Altmannshofer:2015sma,Hurth:2016fbr} but finds a higher significance for a NP contribution.

Some of us discussed in Ref.~\cite{Capdevila:2016ivx} how, under the assumption of lepton-flavour universality violation, the presence of NP in $b\to s\mu\mu$
can be probed in a clean way via the comparison of $b\to s ee$ and $b\to s \mu\mu$ observables, in which hadronic uncertainties cancel. In the present paper we take another approach: we discuss the different sources of hadronic uncertainties and provide robust arguments disfavouring the possibility that these non-perturbative effects are the origin of the observed anomalies. 
At leading order (LO) in the effective Hamiltonian approach, predictions involve two types of contributions, i.e., tree-level diagrams with insertions of the operators 
\begin{eqnarray}
&&{\cal O}_7=\frac{e}{16\pi^2}m_b(\bar{s}\sigma_{\mu\nu}P_Rb)F^{\mu\nu},\nonumber\\
&&{\cal O}_9^{\ell}=\frac{e^2}{16\pi^2}(\bar{s}\gamma_{\mu}P_Lb)(\bar{\ell}\gamma^\mu \ell),\hspace{0.5cm}
{\cal O}_{10}^{\ell}=\frac{e^2}{16\pi^2}(\bar{s}\gamma_{\mu}P_Lb)(\bar{\ell}\gamma^\mu\gamma_5 \ell)
\end{eqnarray}
(generated at one loop in the SM), as well as one-loop diagrams with an insertion of the charged-current operator 
\begin{equation}
{\cal O}_2 = (\bar{s}\gamma^\mu P_L c)(\bar{c}\gamma_\mu  P_L b)
\end{equation}
(generated at tree level in the SM). In contributions of the first type, the leptonic and the hadronic currents factorise, and QCD corrections are restricted to the hadronic $B \to M$ current. This class of {\it factorisable QCD corrections} thus forms part of the hadronic form factors parametrising the $B \to M$ transition. Contributions of the second type, on the other hand, receive {\it non-factorisable QCD corrections} that cannot be absorbed into form factors. 

Both types of corrections have to be taken into account to assess hadronic uncertainties in the computation of $B\to K^*\ell^+\ell^-$ observables. 
In this paper, we collect arguments to demonstrate that the hadronic uncertainties are sufficiently under control 
and we further present counter-arguments to recent articles claiming SM explanations based on incomplete analyses.
In Sec.~\ref{sec:overview}, we recall the main elements of the computation of $B\to K^*\ell^+\ell^-$ and explain our treatment of the various sources of uncertainties,
applied, e.g., in the global fit in Ref.~\cite{Descotes-Genon:2015uva}. In Sec.~\ref{sec:facPC},
we then discuss, in a pedagogical way, the issue of scheme dependence of the {\it factorisable power corrections} of order $\mathcal{O}(\Lambda_{\textrm{QCD}}/m_B)$, 
which was pointed out for the first time in~\cite{Descotes-Genon:2014uoa}. 
We derive explicit formulae for the contribution from {\it factorisable power corrections} to the most important 
observables $P_5^\prime$, $P_2$ and $P_1$, which allow us to confirm in an analytic way the numerical findings of Ref.~\cite{Descotes-Genon:2014uoa}. 
Moreover, we extract the amount of power corrections (including errors) contained in the form factors from Ref.~\cite{Straub:2015ica} 
and find them to be small, typically at the order of $10\%$, in agreement with dimensional arguments.
In Sec.~\ref{sec:nonfacPC}, the role of $c\bar{c}$ loops for {\it non-factorisable QCD corrections} is discussed. In the framework of the effective Hamiltonian, these corrections correspond to a one-loop contribution from the operator ${\cal O}_2$, which can be recast as a contribution to $C_9$ depending on the 
squared dilepton invariant mass $q^2$, the transversity amplitudes $A_j^{L,R}$ ($j=0,\perp,||$) and the hadronic states 
(as opposed to a universal contribution from New Physics).
This $c\bar{c}$ contribution always accompanies the perturbative SM contribution $C_{9\, \rm pert}^{\rm eff\, SM}$ and the NP one $C_9^{\rm NP}$:
\begin{equation} \label{c9eff} C^{{\rm eff} \, B\to K^*}_{9\, j }=C_{9\,\rm pert}^{\rm eff\, SM} + C_9^{\rm NP} + C^{ c \bar c \, {B\to K^*} }_{9\, j}(q^2). \end{equation}
Using a polynomial parametrisation, we performed fits for $C_{9\, j}^{{c\bar c} \, B \to K^*}(q^2)$  in various scenarios.
We discuss the quality of the fits and compare our results with those presented in recent articles~\cite{Ciuchini:2015qxb,Ciuchini:2016weo}, with an emphasis on their statistical interpretation.
In Sec.~\ref{sec:tests}, we provide further experimental tests of hadronic uncertainties, 
in particular $P_2$ at very low $q^2$ that exhibits a kinematic protection with respect to 
charm-loop contributions entering $C_9$, opening the door to a theoretically clean exploration of NP contributions to the Wilson coefficient $C_{10}$. 
We also discuss the recent measurement of the observable $Q_5$ by Belle in terms of NP and SM alternatives.
Sec.~\ref{sec:concl} finally contains our conclusion, while App.~\ref{App:A} provides a dedicated comparison of the parametric uncertainties arising in recent theoretical predictions of $B\to K^*\ell\ell$ observables by different groups and App.~\ref{App:B} our predictions for the observable $R_{K^*}$ in several benchmark scenarios.

\section{An overview of the computation of $B\to K^*\ell^+\ell^-$ observables}\label{sec:overview}

The theoretical framework used in Refs.~\cite{Descotes-Genon:2014uoa,Descotes-Genon:2015uva} to describe the decay $B\to K^*\ell^+\ell^-$ at low 
squared invariant dilepton masses $q^2$ (where the most significant tensions with the SM were found) is based on QCD factorisation  (QCDF) supplemented by a sophisticated estimate of the power corrections of order $\Lambda_{\textrm{QCD}}/m_B$ (\textit{improved} QCDF). The use of effective theories~\cite{Charles:1998dr,Beneke:2000wa,Beneke:2001at,Beneke:2002ph} allows one to relate the different $B\to K^*$ form factors at leading order in $\Lambda_{\textrm{QCD}}/m_B$ and $\Lambda_{\textrm{QCD}}/E$, where $E$ is the energy of the $K^*$. This procedure reduces the required hadronic input from seven to two independent form factors, the so-called \textit{soft form factors} $\xi_\perp,\xi_\|$, which in the region of 
low $q^2$ can be calculated using light-cone sum rules (LCSR). Two sets of LCSR form factors are available in the literature which have been
calculated with very different approaches: KMPW form 
factors~\cite{Khodjamirian:2010vf} that were computed using $B$-meson distribution amplitudes, and BSZ form factors~\cite{Straub:2015ica} that make use of light-meson distribution amplitudes and a prevalent application of equations of motion. The better knowledge of $K^*$-meson distribution amplitudes led to results with a smaller uncertainty in the BSZ case compared to the KMPW computation.
In Refs.~\cite{Descotes-Genon:2014uoa,Descotes-Genon:2015uva} we took advantage of the possibility of comparing the results for the two different sets of form factors as a robustness test of the optimized observables $P^{(\prime)}_i$~\cite{Matias:2012xw,Descotes-Genon:2013vna}. For our default predictions we relied on the KMPW form factors 
which have larger uncertainties and thus lead to more conservative predictions for observables.
By construction the choice of the set of form factors has a relatively low impact on optimized observables but it has a large impact on the error size of 
form-factor sensitive observables like the longitudinal polarisation $F_L$ or the CP-averaged angular coefficients $S_i$.

The large-recoil symmetry limit is enlightening as it allows us to understand the main behaviour of optimized observables in presence of New Physics  in a form-factor independent way. However, for precise predictions of these observables it has to be complemented  with different kinds of corrections, separated in two classes: factorisable and non-factorisable corrections. Improved QCDF~\footnote{``Improved'' stands for ${\cal O}(\Lambda_{\rm QCD}/m_B)$ corrections that go beyond QCDF and are included as uncertainty estimates in our predictions.} provides a systematic formalism to include the different corrections as a decomposition of the amplitude in the following form \cite{Beneke:2001at}:
\begin{equation} \label{amp}
\langle \ell^+\ell^- {\bar K}^*_i | H_{\rm eff} | \bar{B} \rangle= \sum_{a,\pm} C_{i,a} \xi_{a} + \Phi_{B,\pm} \otimes T_{i,a,\pm} \otimes \Phi_{K^*,a} + {\cal O}(\Lambda_{\rm QCD}/m_B).
\end{equation}
Here, $C_{i,a}$ and $T_{i,a}$ ($a=\perp,\|$) are perturbatively computable contributions for
the various $K^*$ polarisations ($i=0,\perp,\|$) 
and $\Phi_{B,K^*}$ denote the light-cone distribution amplitudes of the $B$- and $K^*$-mesons.

\begin{itemize}
\item \textit{Factorisable corrections} are the corrections that can be absorbed into the (full) form factors $F$ by means 
of a redefinition at higher orders in $\alpha_s$ and $\Lambda_{\textrm{QCD}}/m_B$:
\begin{equation}
  F(q^2)\;=\;F^\infty(\xi_\perp(q^2),\xi_\|(q^2))\,+\,\Delta F^{\alpha_s}(q^2)\,+\,\Delta F^\Lambda(q^2).
  \label{eq:FFnlo}
\end{equation}
The two types of corrections to the leading-order form factor $F^{\infty}(\xi_\perp,\xi_\|)$ are
 factorisable $\alpha_s$-corrections $\Delta F^{\alpha_s}$ and factorisable ${\cal O}(\Lambda_{\rm QCD}/m_B)$ corrections $\Delta F^{\Lambda}$. 
While the former can be computed within QCDF and are related to the prefactors $C_{a,i}$ of Eq.~(\ref{amp}), the latter, which 
can be parameterized as an expansion in $q^2/m_B^2$, represent part of the ${\cal O}(\Lambda_{\rm QCD}/m_b)$ terms of Eq.~(\ref{amp}) that QCDF cannot predict. 
In our approach we obtain central values for the $\Delta F^{\Lambda}$ corrections by performing a fit  to the full LCSR form factors $F^{\textrm{LCSR}}$, 
yielding results of typically $(5-10)\%\times F^{\textrm{LCSR}}$ in size, as expected for ${\cal O}(\Lambda_{\rm QCD}/m_B)$  corrections. 
The errors associated to $\Delta F^{\Lambda}$ are estimated by 
varying $\Delta F^{\Lambda}$ in an uncorrelated way in the range of $10\%\times F^{\textrm{LCSR}}$ around the central values. Even though there is no rigorous way in validating this assumption on the error size of power corrections,
we have already shown in Ref.~\cite{Descotes-Genon:2014uoa} that our error assignment of $10\%$ for power corrections is conservative with respect to 
the central values of KMPW form factors, and we show that the same applies for the BSZ form factors~\cite{Straub:2015ica} (including uncertainties). In Sec.~\ref{sec:facPC} we will further discuss the dependence of improved QCDF predictions on the scheme, i.e., on the choice of definition for $\xi_{\perp,\|}$ in terms of full form factors. 
We will argue that
an {\it appropriate scheme} is a scheme that naturally minimizes the sensitivity to power corrections in the relevant observables like $P_5^\prime$.

\item \textit{Non-factorisable corrections} refer to corrections that cannot be absorbed into the definition of the form factors
due to their different structure. One can identify two types of such corrections.
On one side, non-factorisable $\alpha_s$-corrections originating from hard-gluon exchange in diagrams with insertions of 
four-quark operators $\mathcal{O}_{1-6}$ and the chromomagnetic operator $\mathcal{O}_8$: they can be calculated in QCDF \cite{Beneke:2001at} and contribute to $T_{a,i}$ in Eq.~(\ref{amp}). On the other side, there are non-factorisable power corrections of ${\cal O}(\Lambda_{\textrm{QCD}}/m_b)$, 
some of them involving $c\bar{c}$ loops~\footnote{Contributions that do not involve $c\bar{c}$ loops are less important in practice. They will be treated according to the approach described in Sec.~4 of Ref.~\cite{Descotes-Genon:2014uoa}.}. 
The long-distance $c\bar{c}$-loop contribution is included as an additional uncertainty, estimated on the basis of the only existing computation~ \cite{Khodjamirian:2010vf} of soft-gluon emission from four-quark operators involving $c\bar{c}$ currents. 
The calculation in Ref.~\cite{Khodjamirian:2010vf} was done in the framework of LCSRs with $B$-meson distribution amplitudes and makes use of
an hadronic dispersion relation to obtain results in the whole large-recoil region. Taken at face value, the resulting correction would increase the anomaly~\cite{Descotes-Genon:2013wba}. However in our predictions of observables, we add the corresponding corrections to the three transversity amplitudes with prefactors $s_i$ that are scanned from $-1$ to +1:
\begin{equation}
C_9^{{c} \bar{c} \, i}(q^2)= s_i C^{c \bar{c} \, i}_{9\,\rm KMPW}.
\end{equation}
 In this way we allow for the possibility that a large relative phase could flip the sign 
of the long-distance charm contribution~\cite{Descotes-Genon:2015uva}. We note that our conservative approach typically leads to larger uncertainties for observables as compared to 
other estimates in the literature~\cite{Jager:2012uw,Straub:2015ica}. 
\end{itemize}

Finally, it is interesting to notice that the use of a different theoretical approach (full form factors~\cite{Altmannshofer:2008dz}) and of different hadronic input 
(BSZ form factors~\cite{Straub:2015ica}) gives results for the relevant observable $P_5^\prime$ that are in good agreement with ours 
(the predictions agree within $1\sigma$ in every bin). 
In the following, we discuss the impact  of the two types of low-q$^2$ hadronic uncertainties in more detail:
factorisable power corrections (Sec.~\ref{sec:facPC}) and long-distance $c\bar{c}$ loops (Sec.~\ref{sec:nonfacPC}).

\section{Anatomy of factorisable power corrections}\label{sec:facPC}

In the region of large recoil of the $K^*$ meson, the non-perturbative form factors needed for the prediction of $B\to K^*\mu^+\mu^-$ are available from two different LCSR calculations in Refs.~\cite{Khodjamirian:2010vf} (KMPW) and \cite{Straub:2015ica} (BSZ).
In Ref.~\cite{Straub:2015ica}, the set of form factors has been provided together with the corresponding correlations, essential for the cancellation of the form factors at LO in optimized observables.  
Instead of using the results provided in Ref.~\cite{Straub:2015ica}, \emph{the dominant correlations can alternatively be assessed from first principles, by means of large-recoil symmetries} which relate the seven form factors among each other. Among the advantages of this second method, the correlations are free from the model assumptions entering the particular LCSR calculation and \emph{the method can be applied also to sets of form factors for which the correlations have not been specified}, e.g., Ref.~\cite{Khodjamirian:2010vf}. As a drawback, these correlations are obtained only at leading order, and symmetry-breaking corrections of order $\mathcal{O}(\Lambda/m_B)$ have to be estimated from dimensional arguments, implying a scheme dependence of the predictions at $\mathcal{O}(\Lambda/m_B)$. We will discuss this scheme dependence in the following.

\subsection{Scheme dependence}
\label{sec:schemedep}

Theoretical predictions for the decay $B\to K^*\ell^+\ell^-$ depend on seven hadronic form factors usually denoted as
$V,A_0,A_1,A_2,T_1,T_2,T_3$. For small invariant  dilepton masses $q^2\ll m_B^2$ (large-recoil limit), and at leading order in $\alpha_s$ and $\Lambda/m_B$, the set of form factors becomes linearly dependent~\cite{Charles:1998dr,Beneke:2000wa,Beneke:2001at,Beneke:2002ph}:
\begin{eqnarray}
   \frac{m_B}{m_B+m_V}V(q^2)\,=\,\frac{m_B+m_V}{2E}A_1(q^2)\,=\,T_1(q^2)\!&=&\!\frac{m_B}{2E}T_2(q^2)\,\equiv\,\xi_{\perp}(E),\nonumber\\
   \frac{m_V}{E}A_0(q^2)\,=\,\frac{m_B+m_V}{2E}A_1(q^2)-\frac{m_B-m_V}{m_B}A_2(q^2)\!&=&\!\frac{m_B}{2E}T_2(q^2)-T_3(q^2)\,\equiv\,\xi_\|(E).
   \label{eq:LaRecRel}
\end{eqnarray}
Here, $m_B$ and $m_{K^*}$ are the meson masses, and $E$ is the energy of the $K^*$.
Within the above-mentioned approximations the number of independent form factors thus reduces to two,
the so-called soft form factors $\xi_\perp$ and $\xi_\|$, and the full set of form factors
$V,A_0,A_1,A_2,T_1,T_2,T_3$ can be obtained as linear combinations of $\xi_\perp$, $\xi_\|$.

Eqs.~(\ref{eq:LaRecRel}) allow us to construct observables in which the form factors cancel at
leading order. For an illustration, let us focus at $q^2=0$, 
where the first relation in Eq.~(\ref{eq:LaRecRel}) implies
\begin{equation}
   \frac{A_1(0)}{T_1(0)}=\frac{T_1(0)}{V(0)}=\frac{V(0)}{A_1(0)}=1+\mathcal{O}(\alpha_s,\Lambda/m_B),
   \label{eq:FFratios}
\end{equation}
while $T_1(0)/T_2(0)=1$ holds exactly due to a kinematic identity from the definition of $T_1$ and $T_2$. Observables involving ratios like the
ones in Eq.~(\ref{eq:FFratios}) are independent of the form factor input up to effects of 
$\mathcal{O}(\alpha_s,\Lambda/m_B)$, and the optimized observables $P^{(\prime)}_i$ are defined following
this philosophy. The reduced sensitivity to the hadronic form factor input renders these observables sensitive to subleading sources of uncertainties, i.e. to effects of $\mathcal{O}(\alpha_s)$ and $\mathcal{O}(\Lambda/m_B)$.
While ${\cal O}(\alpha_s)$ corrections to Eqs.~({\ref{eq:LaRecRel}}) can be included in the framework of QCDF, the so-called factorisable power corrections of $\mathcal{O}(\Lambda/m_B)$ are not computable in QCDF. 

Accurate QCDF predictions rely in an essential way on quantifying the uncertainty due to power-suppressed $\Lambda/m_B$ effects. This is typically done
by assigning uncorrelated errors of the size $\delta\sim 10\%$ to Eq.~({\ref{eq:LaRecRel}}) (and thus
to the ratios in Eq.~(\ref{eq:FFratios})). Note, however, that this cannot be done in a unique way. 
Let us, for instance, assume that the errors on $A_1(0)/T_1(0)$ and $T_1(0)/V(0)$ are given by $\delta_1$ and $\delta_2$, respectively:
\begin{equation}
   \frac{A_1(0)}{T_1(0)}=1\pm \delta_1,\hspace{2cm}
   \frac{T_1(0)}{V(0)}=1\pm \delta_2.
   \label{eq:RatErrs1}
\end{equation}
The error $\delta_3$ on the ratio $A_1(0)/V(0)$ is then fixed by
\begin{equation}
   1\pm\delta_3=\frac{A_1(0)}{V(0)}=\frac{A_1(0)}{T_1(0)}\frac{T_1(0)}{V(0)}=
   \left\{\begin{array}{l} 1\pm\sqrt{\delta_1^2+\delta_2^2}, \hspace{0.5cm}\textrm{quadratic 
error propagation}\\[1ex]
1\pm(\delta_1+\delta_2), \hspace{0.7cm}\textrm{linear error propagation} 
          \end{array}\right.,
   \label{eq:RatErrs2}
\end{equation}
depending on how uncertainties are propagated.
The assumption of a universal error size $\delta_1=\delta_2\equiv\delta$
for the first two ratios thus leads to an error $\delta_3=\sqrt{2}\delta$ or $\delta_3=2\delta$ 
for the third one, although in principle the three ratios should be treated on an equal footing.

The same phenomenon can be understood also from a different point of view. In the QCDF approach, predictions of 
observables depend on the two soft form factors $\xi_\perp$ and $\xi_\|$ for which hadronic input 
(from LCSR) is needed. 
According to Eq.~({\ref{eq:LaRecRel}}), there are various possibilities 
to select the input among the seven full factors $V,A_1,A_2,A_0,T_1,T_2,T_3$, and the choice defines an input scheme.
One possible choice would consist for example in defining 
\begin{eqnarray}
  \xi_\perp(q^2)&=&\frac{m_B}{m_B+m_V}V(q^2),\nonumber\\
  \xi_\|(q^2)&=&\frac{m_B+m_V}{2E}A_1(q^2)-\frac{m_B-m_V}{m_B}A_2(q^2)\hspace{1.5cm}
  \textrm{(scheme 1).}
  \label{eq:scheme1}
\end{eqnarray} 
A different choice would consist in identifying 
\begin{equation}
  \xi_\perp(q^2)\,=\,T_1(q^2),\hspace{1.5cm}
  \xi_\|(q^2)\,=\,\frac{m_V}{E}A_0(q^2)\hspace{2.3cm}
  \textrm{(scheme 2).}
  \label{eq:scheme2}
\end{equation}
By definition, the form factors (or linear combinations of form factors) taken as input are exactly known
to all orders in $\alpha_s$ and $\Lambda/m_B$. The remaining form factors are then determined from 
the symmetry relations in Eq.~({\ref{eq:LaRecRel}}) upon including $\mathcal{O}(\alpha_s)$ corrections via QCDF and 
assigning an error estimate to unknown $\mathcal{O}(\Lambda/m_B)$ corrections. 
Taking, for instance, as in scheme 2, $T_1(0)=T_1^{\textrm{LCSR}}(0)$ as input for $\xi_\perp(0)$ leads to
\begin{equation}
   V(0)=T_1^{\textrm{LCSR}}(0)+a_V^{\alpha_s}+a_V^\Lambda+...,\hspace{2cm}
   A_1(0)=T_1^{\textrm{LCSR}}(0)+a_{A_1}^{\alpha_s}+a_{A_1}^\Lambda+...,
   \label{eq:PertSeries}
\end{equation}
where $a_V^{\alpha_s},a_{A_1}^{\alpha_s}$ and  $a_V^{\Lambda},a_{A_1}^{\Lambda}$ 
are $\alpha_s$ and $\Lambda/m_B$ corrections to Eq.~({\ref{eq:LaRecRel}}) for each form factor and the ellipsis represents
terms of higher orders. If Eq.~(\ref{eq:PertSeries}) was determined to all orders in $\alpha_s$ and 
$\Lambda/m_B$, predictions for observables would not depend on the chosen input scheme. 
In practice, QCD corrections are known in QCDF up to $\mathcal{O}(\alpha_s^2)$~\cite{Beneke:2005gs,Bell:2010mg} while  $\Lambda/m_B$ corrections 
can only be estimated, implying a scheme dependence in the computation of the observables at $\mathcal{O}(\Lambda/m_B)$ and 
$\mathcal{O}(\alpha_s^3)$.

While the form factors taken as input inherit their uncertainties directly from the LCSR calculation,
the remaining form factors receive an additional error for the unknown $\Lambda/m_B$ corrections $a^\Lambda$. 
In the example above (scheme 2), we have
\begin{equation}
  T_1(0)=T_1^{\textrm{LCSR}}(0)\pm\Delta T_1^{\textrm{LCSR}}(0),
\end{equation}
with $\Delta T_1^{\textrm{LCSR}}(0)$ denoting the uncertainty of the LCSR calculation, and
\begin{eqnarray}
   V(0)&=&(T_1^{\textrm{LCSR}}(0)+a_V^{\alpha_s}+a_V^{\Lambda})\,\pm\,(\Delta T_1^{\textrm{LCSR}}(0) + {\Delta a_V^{\alpha_s}} + \Delta a_V^\Lambda),\nn\\
   A_1(0)&=&(T_1^{\textrm{LCSR}}(0)+a_{A_1}^{\alpha_s}+a_{A_1}^{\Lambda})\,\pm\,
            (\Delta T_1^{\textrm{LCSR}}(0) +{\Delta a_{A_1}^{\alpha_s}} + \Delta a_{A_1}^\Lambda).
\end{eqnarray}
In this case, $V(0)$ and $A_1(0)$ are subject to two main sources of uncertainties, 
namely the error $\Delta T_1^{\textrm{LCSR}}(0)$ of the LCSR calculation and the uncertainties $\Delta a_{V,A_1}^\Lambda$
from unknown power corrections (we neglect the uncertainty $\Delta a_{V,A_1}^{\alpha_s}$ from the perturbative contribution). On the other hand, if we had chosen $V(0)$ or $A_1(0)$ directly as input
for the soft form factor $\xi_\perp(0)$, the only source of error for $V(0)$ or $A_1(0)$ would have been
the respective LCSR error $\Delta V^{\textrm{LCSR}}(0)$ or $\Delta A_1^{\textrm{LCSR}}(0)$. The choice of scheme
thus defines the precision to which the various full form factors are known, keeping those taken as input free
from a pollution by power corrections.

The freedom to choose between different input schemes is equivalent to the ambiguity in implementing
the $10\%$ requirement on the symmetry-breaking corrections to Eqs.~({\ref{eq:LaRecRel}}) and (\ref{eq:FFratios}).
In the scheme 2, the uncertainties on the form factor ratios are:
\begin{eqnarray}
   \frac{A_1(0)}{T_1(0)}&=&1\pm\dfrac{\Delta a_{A_1}^\Lambda}{T_1^{\text{LCSR}}},\hspace{1cm}
   \frac{T_1(0)}{V(0)}\;=\;1\pm\dfrac{\Delta a_{V}^\Lambda}{T_1^{\text{LCSR}}},\nn\\
   \frac{A_1(0)}{V(0)}&=&
   \left\{\begin{array}{l} 1\pm\sqrt{\left(\dfrac{\Delta a_{A_1}^\Lambda}{T_1^{\text{LCSR}}}\right)^2+\left(\dfrac{\Delta a_{V}^\Lambda}{T_1^{\text{LCSR}}}\right)^2}, 
                           \hspace{0.5cm}\textrm{quadratic error propagation}\\[3ex]
                           1\pm\left(\dfrac{\Delta a_{A_1}^\Lambda}{T_1^{\text{LCSR}}}+\dfrac{\Delta a_{V}^\Lambda}{T_1^{\text{LCSR}}}\right), \hspace{1.8cm}\textrm{linear error propagation} 
          \end{array}\right..
   \label{eq:RatErrs3}
\end{eqnarray}
In this expressions we have kept only the errors of $\mathcal{O}(\Lambda/m_B)$ and we have neglected
uncertainties suppressed by additional powers of $\alpha_s$ or $\Lambda/m_B$. Note that the LCSR error $\Delta T_1^{\textrm{LCSR}}(0)$
cancels in this approximation.
Identifying $\delta_1=\Delta a_{A_1}^\Lambda/T_1^{\text{LCSR}}$ and $\delta_2=\Delta a_{V}^\Lambda/T_1^{\text{LCSR}}$, we find that the resulting errors are in agreement with Eqs.~(\ref{eq:RatErrs1}) and (\ref{eq:RatErrs2}). 

How can  the ambiguity from the scheme dependence be solved? To answer this question, let us 
first have a look at the decay $B\to K^*\gamma$. The prediction of this branching ratio depends on the single form factor $T_1(0)$
and the natural choice thus consists in taking its LCSR value directly as input for the theory predictions~\footnote{This decay also receives a contribution from charm loops. For the sake of the argument presented in this section, we will neglect this effect, which should however be included in an actual computation of this branching ratio, contrary to the approach of Ref.~\cite{Jager:2012uw}. We will include this contribution when discussing the fits to $c\bar{c}$ contributions, see Sec.~\ref{sec:ccfit} and in particular Tab.~\ref{tab:B}.}. 
Of course, one could take as input any other form factor to which $T_1$ is related through the symmetry relations in Eq.~(\ref{eq:LaRecRel}), e.g. $V$.
Unlike $T_1$, the choice of $V$ would generate power corrections of $\mathcal{O}(\Lambda/m_B)$ in the prediction for 
$B\to K^*\gamma$, reflecting the fact that the identification $V=T_1$ is only an approximation, valid up to 
$\mathcal{O}(\Lambda/m_B)$, and that the ``wrong'' form factor, $V$, has been used for the prediction
instead of the ``correct'' one, $T_1$. The corresponding increase in the uncertainties is thus caused artificially by an {\it inappropriate} choice of the input scheme. This becomes even more obvious in the hypothetical limit where the errors of the LCSR calculation go to zero: In this case, the prediction for $B\to K^*\gamma$ would be free from any form factor uncertainty (as it should be) when $T_1$ is taken as input, while the wrong central value would be obtained when $V$ is used, together with an irreducible error of order $\mathcal{O}(|V^{\text{LCSR}}-T_1^{\text{LCSR}}|)$.

The example of $B\to K^*\gamma$ clearly illustrates the fact that an inappropriate choice of scheme can artificially 
increase the uncertainty of the theory prediction. The situation is less obvious in the case of $B\to K^*\mu^+\mu^-$, where typically all seven form factors enter the prediction of the observables. Ignoring the form factor $A_0$, whose
contribution is  suppressed by the lepton mass, we observe that the form factors $V,A_1,A_2$ enter the amplitudes together with the Wilson coefficients $C_{9,10}^{(\prime)}$, whereas $T_1,T_2,T_3$ enter the amplitudes together with the coefficient $C_7^{(\prime)}$. In the SM, $C_7^{\text{eff}}\ll \text{Re}(C_9^{\text{eff}})$ (where the effective coefficients $C_{7,9}^{\text{eff}}$ include effects from
perturbative $q\bar{q}$ loops), e.g. $C_7^{\text{eff}}(q_0^2)=-0.29$ and $\text{Re}(C_9^{\text{eff}})(q_0^2)=4.7$ at $q^2_0=6$\,GeV$^2$.
Hence the (axial-)vector form factors $V,A_1,A_2$ are in general more relevant than the tensor form factors $T_1,T_2,T_3$, except for the very low $q^2$-region where the $C_7$ contribution can be enhanced by the $1/q^2$ pole from the photon propagator. In particular in the anomalous bins of the observable $P_5^\prime$ ($4\leq q^2\leq 8$ GeV$^2$), we find that the impact from $C_7$ is strongly suppressed compared to the 
impact from $C_9$. This can be seen by setting some of the Wilson coefficients to zero and determining the resulting change in the predictions: one gets a shift of $\Delta P_5^\prime(C_7=0)_{[4,6]}=-0.19$ when $C_7$ is switched off, compared to $\Delta P_5^\prime(C_9=0)_{[4,6]}=+1.34$ when $C_9$ is switched off. With respect to the soft form factor $\xi_\perp$, the observable $P_5^\prime$ is thus dominated by the ratio $A_1/V$ suggesting the form factor $V$, or alternatively $A_1$, as a natural input for $\xi_\perp$. Defining $\xi_\perp$ from $T_1$, as done in 
Refs.~\cite{Jager:2012uw,Jager:2014rwa}, on the other hand represents an inadequate choice: to a good approximation, the prediction of $P_5^\prime$ in the anomalous bins
does not depend on this form factor, due to a suppression by $|C_7/C_9|\ll 1$. 

Together with the linear propagation of errors applied in Refs.~\cite{Jager:2012uw,Jager:2014rwa}, the choice 
of $T_1$ as input leads to an artificial inflation of the uncertainty by a factor of $2$
in the anomalous bins of $P_5^\prime$, as we demonstrated in Eqs.~(\ref{eq:RatErrs2}) and (\ref{eq:RatErrs3}). In other words, we conclude that the results on $P_5^\prime$ obtained in Ref.~\cite{Jager:2014rwa} correspond to an implicit assumption of $20\%$ power 
corrections~\footnote{This is in contradiction with the assumption initially stated in Ref.~\cite{Jager:2014rwa} that a 10\% power correction is used for all the form factors.} because this is the size of symmetry breaking implicitly assumed for the dominant form factor ratio $A_1/V$~\footnote{This provides only a partial explanation to the larger uncertainties in Ref.~\cite{Jager:2012uw}. Apart from a factor of two in the error assigned to factorisable power corrections that we have just discussed, Ref.~\cite{Jager:2012uw} also states much larger
parametric errors compared to Ref.~\cite{Descotes-Genon:2014uoa} and Refs.~\cite{Altmannshofer:2014rta,Altmannshofer:2015sma}.  This is surprising, given the fact that the 
uncertainties assumed for the key parameters like $m_c$ are compatible, while the errors for the 
form factors are even significantly smaller in Ref.~\cite{Jager:2012uw} due to the extraction of $T_1(0)$ using experimental data (see also App.~\ref{App:A}).}. The situation is different for observables that vanish in the limit $C_7\to 0$, i.e. that depend on $C_7$ already
at leading order in $C_7/C_9$, like the observable $P_2$. In this case, it is not clear a priori whether the observable is more sensitive to the (axial-)vector or to the tensor form factors, and the answer to this question
requires a closer inspection (see Sec.~\ref{sec:PowCorrFormulae}).

In summary, in the soft-form factor approach, we expect the uncertainties of our predictions to be scheme dependent. An inappropriate choice of definition for the soft form factors will inflate the errors on the predictions. For each observable, we should thus choose a scheme as appropriate as possible to avoid an overestimation of the uncertainties.

\begin{table}
\footnotesize
\centering
\begin{tabular}{@{}l|rrr|ccc@{}}
  & $a_F\qquad$ & $b_F\qquad$ & $c_F\qquad$ &
$r(0\,\rm{GeV}^2)$ & $r(4\,\rm{GeV}^2)$ & $r(8\,\rm{GeV}^2)$ \\ 
\hline 
$A_0$& $0.000\pm 0.000$ & $0.054\pm 0.033$ & $0.197\pm 0.203$ 
     & $0.000\pm 0.000$ & $0.026\pm 0.020$ & $0.055\pm 0.047$ \\
     & $\phantom{0.000}\pm 0.000$ & $\phantom{0.040}\pm 0.054$ & $\phantom{0.177}\pm 0.112$ 
     & $\phantom{0.000}\pm 0.000$ & $\phantom{0.019}\pm 0.020$ & $\phantom{0.041}\pm 0.038$ \\ 
$A_1$& $0.020\pm 0.011$ & $0.036\pm 0.025$ & $0.037\pm 0.049$ 
     & $0.071\pm 0.043$ & $0.086\pm 0.045$ & $0.102\pm 0.054$ \\  
     & $\phantom{0.022}\pm 0.029$ & $\phantom{0.045}\pm 0.017$ & $\phantom{0.102}\pm 0.022$ 
     & $\phantom{0.076}\pm 0.100$ & $\phantom{0.096}\pm 0.100$ & $\phantom{0.121}\pm 0.100$ \\ 
$A_2$& $0.028\pm 0.016$ & $0.079\pm 0.038$ & $0.131\pm 0.079$ 
     & $0.116\pm 0.070$ & $0.147\pm 0.078$ & $0.188\pm 0.099$ \\
     & $\phantom{0.031}\pm 0.041$ & $\phantom{0.095}\pm 0.048$ & $\phantom{0.239}\pm 0.056$ 
     & $\phantom{0.125}\pm 0.165$ & $\phantom{0.165}\pm 0.174$ & $\phantom{0.210}\pm 0.182$ \\
$T_1$& $-0.017\pm 0.013$ & $-0.017\pm 0.009$ & $-0.037\pm 0.023$ 
     & $0.061\pm 0.045$ & $0.057\pm 0.038$ & $0.054\pm 0.030$ \\  
     & $\phantom{-0.011}\pm 0.031$ & $\phantom{-0.005}\pm 0.043$ & $\phantom{-0.015}\pm 0.090$ 
     & $\phantom{0.037}\pm 0.100$ & $\phantom{0.032}\pm 0.100$ & $\phantom{0.027}\pm 0.100$ \\  
$T_2$& $-0.017\pm 0.012$ & $0.007\pm 0.027$ & $0.025\pm 0.053$ 
     & $0.061\pm 0.045$ & $0.050\pm 0.045$ & $0.036\pm 0.053$ \\ 
     & $\phantom{-0.011}\pm 0.031$ & $\phantom{0.014}\pm 0.016$ & $\phantom{0.139}\pm 0.027$ 
     & $\phantom{0.037}\pm 0.100$ & $\phantom{0.019}\pm 0.100$ & $\phantom{0.011}\pm 0.100$ \\ 
$T_3$& $-0.007\pm 0.021$ & $0.014\pm 0.041$ & $0.061\pm 0.208$ 
     & $0.037\pm 0.111$ & $0.013\pm 0.132$ & $0.016\pm 0.176$ \\ 
     & $\phantom{-0.011}\pm 0.018$ & $\phantom{-0.007}\pm 0.019$ & $\phantom{0.102}\pm 0.026$ 
     & $\phantom{0.060}\pm 0.100$ & $\phantom{0.046}\pm 0.100$ & $\phantom{0.018}\pm 0.100$ \\ 

\end{tabular}
\caption{Results for the fit of the power-correction parameters $a_F,b_F,c_F$ 
to the $B\to K^*$ form factors from Ref.~\cite{Straub:2015ica}, using the input scheme 1
in the transversity basis.
Furthermore, the relative size $r(q^2)$ with which the power corrections contribute to the full form factors 
is shown for $q^2=0,4,8\,\rm{GeV}^2$. In the first line of each entry, the central value and 
the error obtained from the fit are given. In the second line, the estimate $\Delta F^\Lambda=10\%\times F^{\rm LCSR}$ is displayed for comparison. }
\label{tab:fitPC}
\end{table}

\subsection{Correlated fit of power corrections to form factors}
\label{sec:BSZfit}

Having clarified the issue of the scheme dependence, we can turn to
the question of the actual size $\delta$ of the symmetry breaking corrections. 
Both Refs.~\cite{Descotes-Genon:2014uoa} and \cite{Jager:2014rwa} use $\delta=10\%$ as an error estimate.
It is instructive to study how this ad-hoc value compares to the size 
of power corrections present in specific LCSR calculations.
In Ref.~\cite{Descotes-Genon:2014uoa}, we extracted information on power corrections
from the LCSR form factors in
Refs.~\cite{Khodjamirian:2010vf} (KMPW) and \cite{Ball:2004rg} (BZ), and we will discuss the results from Ref.~\cite{Straub:2015ica} in a similar way, checking the robustness of this extraction.

The form factors $F$ are parametrised according to Eq.~(\ref{eq:FFnlo}).
For a specific set of LCSR form factors 
$\{F^\text{LCSR}(q^2)\}$, the power corrections $\Delta F^\Lambda(q^2)$ can then be determined as the 
difference between the full $F^{\text{LCSR}}(q^2)$ and the large-recoil result 
$F^\infty(q^2)$ upon including $\alpha_s$-corrections $\Delta F^{\alpha_s}(q^2)$ from QCDF.
In practice we fit the coefficients $a_F,b_F,c_F$ of the parametrisation
\begin{equation}\label{eq:PCparaJager}
   \Delta F^{\Lambda}(q^2) = a_F\,+\,b_F\,\frac{q^2}{m_B^2}\,+\,c_F\,\frac{q^4}{m_B^4}\,+\,\ldots\;.
\end{equation}
to the central value of the LCSR results.
In Tab.~\ref{tab:fitPC} we show the results obtained
within this approach initiated in Ref.~\cite{Descotes-Genon:2014uoa} and applied now to the form factors from Ref.~\cite{Straub:2015ica}. 

In contrast to previous LCSR calculations, Ref.~\cite{Straub:2015ica} for the first time provided the  correlations among the form factors, enabling us to fit
not only the central values of the parameters $a_F,b_F,c_F$ but also their uncertainties according to the correlation matrix of the form factors, 
which will serve us to illustrate the good control of our method of factorisable power corrections.
Tab.~\ref{tab:fitPC} displays the results for the input scheme 1,
defined in Eq.~(\ref{eq:scheme1}), and parametrising power corrections in the transversity basis
$\{V,A_1,A_2,A_0,T_1,T_2,T_3\}$ (this corresponds to the default choice in Ref.~\cite{Descotes-Genon:2014uoa}).

The relative size of power corrections,
\begin{equation}\label{eq:rdef}
   r(q^2)\,=\,\left|\frac{a_F+b_F\frac{q^2}{m_B^2}+c_F\frac{q^4}{m_B^4}}{F(q^2)}\right|,
\end{equation}
is displayed  on the right-hand side of Tab.~\ref{tab:fitPC}
for different invariant masses $q^2=0\,\rm{GeV}^2$, $4\,\rm{GeV}^2$, $8\,\rm{GeV}^2$ of the lepton pair.
Typically, the central values of the power corrections are within the range of $(5-10)\%$, with uncertainties below $5\%$.
These findings are in line with the results 
for the central values of the form factors from Refs.~\cite{Ball:2004rg} (BZ) and \cite{Khodjamirian:2010vf}  (KMPW) obtained in Ref.~\cite{Descotes-Genon:2014uoa}.
Exceptions occur at large $q^2$ for the form factors $A_2$ and $T_3$, which are calculated as linear combination of two functions in Ref.~\cite{Straub:2015ica}.
In the case of $A_2$, the central values of the power corrections reach up to 19$\%$, while the respective uncertainties still do not
exceed $10\%$. Note that in scheme 1, the power corrections to $A_2$ are not an independent function, but they are fixed 
from the ones to $A_1$ as detailed in Ref.~\cite{Descotes-Genon:2014uoa}. In the case of $T_3$, the central values are quite small but come with uncertainties that grow up to $18\%$. It turns out that the power corrections to these two form factors have no impact on the key observables $P_5^\prime$, $P_1$ and $P_2$ as can be seen from the analytic formulae in Sec.~\ref{sec:PowCorrFormulae}, where these terms are either absent or numerically suppressed.

 \begin{figure}
 \begin{center}
 \includegraphics[width=6cm,height=6cm]{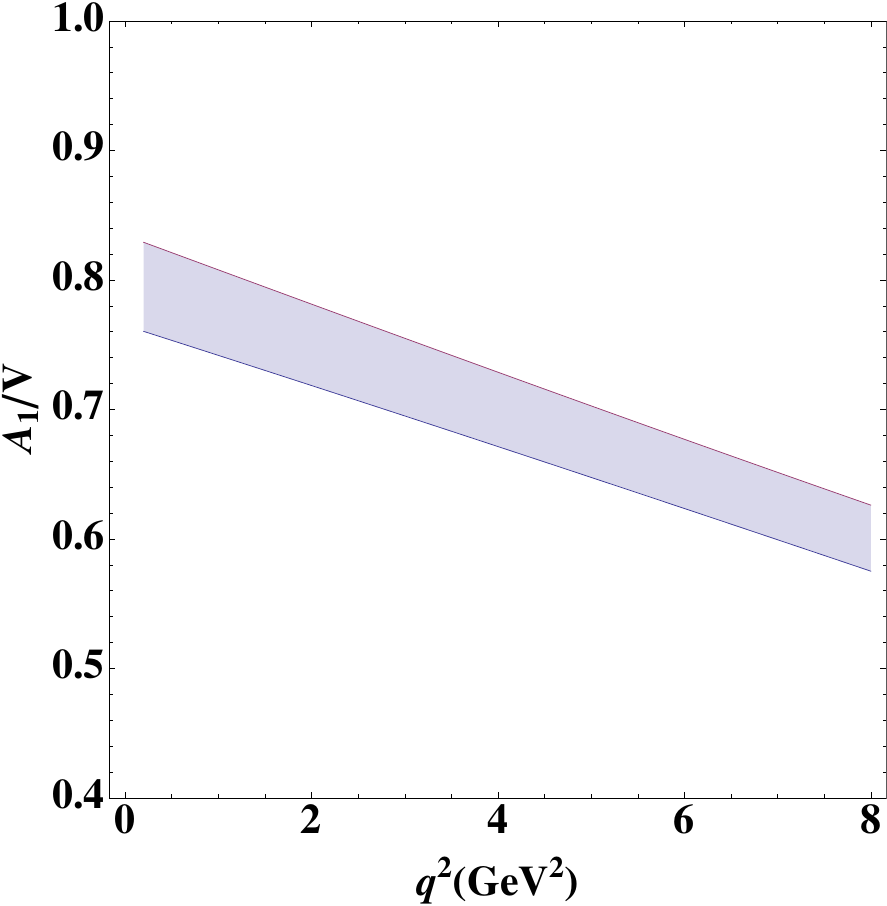}\hspace{1cm}
  \includegraphics[width=6cm,height=6cm]{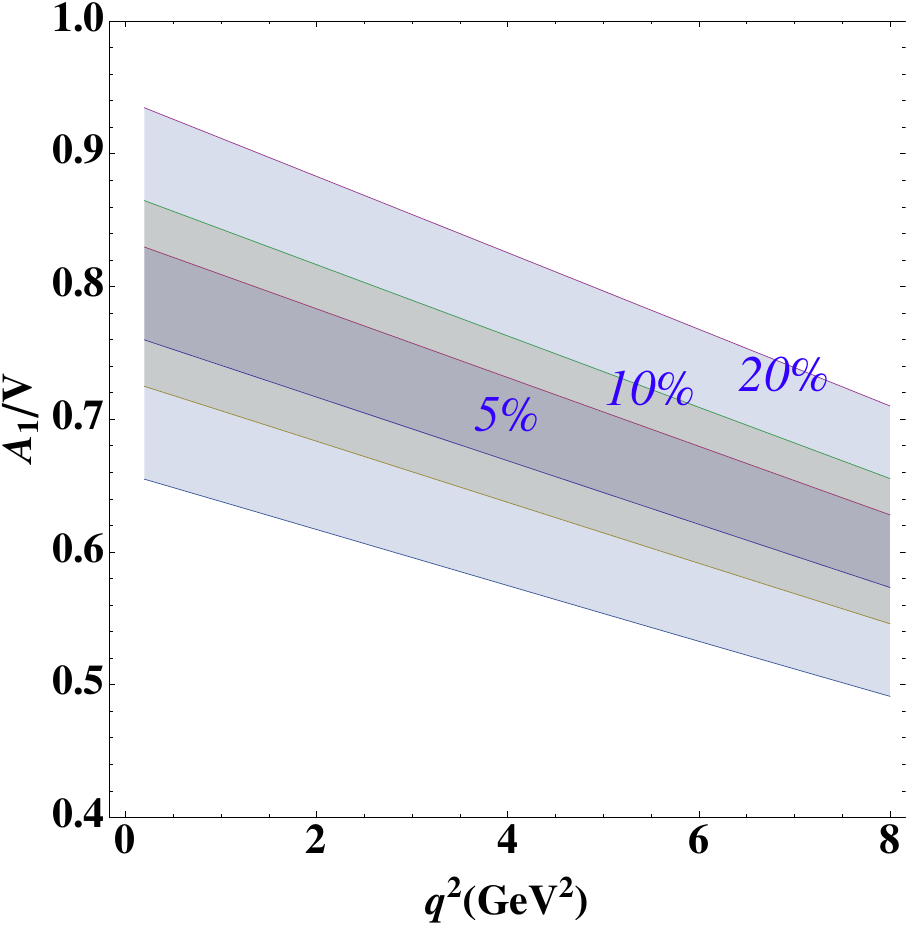} 
  \end{center}
  \caption{Ratio of form factors $A_1/V$ applying the full and soft form factor approaches to the results of Ref.~\cite{Straub:2015ica}. Left: error band according to the LCSR calculation from Ref.~\cite{Straub:2015ica}.
  Right: error bands following the soft form factor approach with $\delta=5\%,10\%,20\%$ power corrections.\label{fig:A1V}}
 \end{figure}

For comparison, Tab.~\ref{tab:fitPC} also features the estimate of power corrections by a generic size of $\delta=10\%$ 
following the approach of Ref.~\cite{Descotes-Genon:2014uoa} to estimate the uncertainties on ${a}_F,{b}_F,{c}_F$ in the absence of information on the correlations among form factors. By definition, the ratio $r(q^2)$ yields $10\%$ for these estimates for all form factors,
except for $A_0$ and $A_2$ where the power corrections are not independent but follow from correlations among form factors.
The comparison with the results from the fit shows that the estimate of power corrections by a generic size of $\delta=10\%$ in Refs.~\cite{Descotes-Genon:2014uoa} is conservative
compared to the procedure followed in Refs.~\cite{Altmannshofer:2014rta,Altmannshofer:2015sma,Hurth:2016fbr,Ciuchini:2015qxb,Ciuchini:2016weo}
consisting in a direct extraction of the errors from the uncertainties given in Ref.~\cite{Straub:2015ica}. 
This is further illustrated in Fig.~{\ref{fig:A1V}}, where the form factor ratio $A_1/V$ dominating the observable $P_5^\prime$ is shown, comparing the direct error assessment from Ref.~\cite{Straub:2015ica} (left plot) and our results from uncertainty assignments of $\delta=5\%,10\%,20\%$
power corrections.

Let us now illustrate how the treatment of power corrections affects the uncertainties
of relevant $B\to K^*\ell\ell$ observables. Taking the above results, and following a similar procedure for the scheme 2 defined in Sec.~\ref{sec:schemedep}, we can compute the SM prediction for $P_5^\prime$ in the anomalous bin $[4,6]$\,GeV$^2$ together with the error from soft form factors and factorisable power corrections (all other sources of errors have been switched off). The results are given in Tab.~\ref{tab:P5p} for the two schemes, with three different options for the treatment of power corrections:
\begin{itemize}
  \item[a)] Estimating the error size of $a_F,b_F,c_F$ as $\sim 10\%\times F^{\rm LCSR}$ and including only 
        the correlations dictated by the large-recoil symmetries. LCSR input is only used 
        to extract the soft form factors $\xi_\perp$ and $\xi_\|$ which are considered as uncorrelated. 
  \item[b)] Determining the errors of $a_F,b_F,c_F$ from the fit to the form factors from Ref.~\cite{Straub:2015ica}
        but including only the correlations dictated by the large-recoil symmetries exactly as in the previous case.
  \item[c)] Determining the errors of $a_F,b_F,c_F$ from a correlated fit to the              
        form factors from Ref.~\cite{Straub:2015ica} and including the correlations between the $a_F,b_F,c_F$ and
        the soft form factors $\xi_\perp,\xi_\|$ as extracted from the correlation matrix in Ref.~\cite{Straub:2015ica}. 
\end{itemize}
\begin{table}
\small
\centering
\renewcommand{\arraystretch}{1.5}
\begin{tabular}{@{}c|c|c@{}}
$\av{P_5^\prime}_{[4.0,6.0]}$ & scheme 1 & scheme 2  \\
\hline
a & $-0.72\pm 0.05$ & $-0.72\pm 0.15$ \\
\hline
b & $-0.72\pm 0.03$ & $-0.72\pm 0.04$ \\
\hline
c & $-0.72\pm 0.03$ & $-0.72\pm 0.03$ \\
\hline
full BSZ& \multicolumn{2}{c}{$-0.72\pm 0.03$} 
\end{tabular}
\caption{SM prediction for $P_5^\prime$ in the anomalous bin $[4,6]$\,GeV$^2$ together with the error from soft form factors and factorisable power corrections (all other sources of errors have been switched off). Results are shown for the three different options for the treatment of power corrections and for the two different input schemes discussed in the text. The last row contains the prediction from a direct use of the full form factors from Ref.~\cite{Straub:2015ica}.}
\label{tab:P5p}
\end{table} 
The error estimate in option a) is mainly based on the fundamental large-recoil symmetries
and thus to a large extent independent of the details of the particular LCSR calculation~\cite{Straub:2015ica}. When going over
option b) to c), we include in each step more information from Ref.~\cite{Straub:2015ica} (the actual size of power corrections for option b), and the correlations for option c)). With option c) the full information from the particular 
LCSR form factors is used, implying that the result must be independent of the input scheme
(apart from a residual scheme dependence from non-factorisable power corrections) and that it must coincide with the one obtained by a direct use of the correlated full form factors (displayed in the last row of Tab.~\ref{tab:P5p}).  
The numerical confirmation of this correspondence provides a consistency check for our implementation of the fit
of the power corrections and the various methods. 

In Tab.~\ref{tab:P5p}, the errors obtained in option b) are very similar to the ones using option c). 
From this observation we conclude that the correlations among the power correction parameters $a_F,b_F,c_F$ 
and the ones among the soft form factors $\xi_\perp,\xi_\|$ have very little impact and that the dominant form factor
correlations are indeed the ones from the large-recoil symmetries. The difference in the errors for option a) between scheme 1 and scheme 2 is easily understood: while the LCSR results of Ref.~\cite{Straub:2015ica} end up with about $\delta\sim 5\%$ power corrections, 
a generic size of $\delta=10\%$ is assumed for option a). In scheme 1, this leads to the expected increase of the errors by roughly a factor 2. On the other hand, in scheme 2,  we find an increase of the errors by more than a factor 4,
in accordance with the discussion in the previous section. As argued there, the implementation of option a)
in scheme 2 actually corresponds to the assumption of $\delta=20\%$ power corrections for the relevant form factor ratio $A_1/V$.

\subsection{Analytic formulae for factorisable power corrections to optimized observables}
\label{sec:PowCorrFormulae}

We have considered a particular observable and demonstrated numerically that the prediction for observables depends on the scheme chosen
for the soft form factors $\xi_{\perp,\|}$. 
In this section we illustrate this scheme dependence more explicitly by giving analytic formulae for the power corrections to the observables
$P_5^\prime$, $P_1$ and $P_2$, both in the transversity and in the helicity basis. The two bases are related to each other via the relations given in Eq.~(31) of Ref.~\cite{Jager:2012uw}. In both cases we parametrize the power corrections according to Eq.~(17). The formulae are 
given without fixing a particular scheme, i.e., before power corrections are partially absorbed into the 
non-perturbative input parameters $\xi_\perp$ and $\xi_\|$.

In the helicity basis, the formula for $P_5^\prime$ reads
\begin{align}
P_5^\prime=P_5^\prime|_\infty\Bigg(1&+\dfrac{2a_{V_-}-2a_{T_-}}{\xi_\perp}
\dfrac{C_7^\text{eff}(C_{9,\perp}C_{9,\parallel}-C_{10}^2)}{(C_{9,\perp}+C_{9,\parallel})(C_{9,\perp}^2+C_{10}^2)}\dfrac{m_b m_B}{q^2}\nonumber\\
&-\dfrac{2a_{V_+}}{\xi_\perp}\dfrac{C_{9,\parallel}}{C_{9,\perp}+C_{9,\parallel}}+\dfrac{2a_{V_0}-2a_{T_0}}{\tilde{\xi}_\parallel}
\dfrac{C_7^\text{eff}(C_{9,\perp}C_{9,\parallel}-C_{10}^2)}{(C_{9,\perp}+C_{9,\parallel})(C_{9,\parallel}^2+C_{10}^2)}\frac{m_b}{m_B}\nonumber\\
&+\text{nonlocal terms}\Bigg)+O\left(\dfrac{m_{K^*}}{m_B},\dfrac{\Lambda^2}{m_B^2},\dfrac{q^2}{m_B^2}\right),\label{eq:analyticP5p}
\end{align}
where $\tilde{\xi}_\parallel=(E_{K^*}/m_{K^*})\, \xi_\parallel$ and  following Ref.~\cite{Jager:2014rwa}, we have defined 
\begin{equation}
C_{9,\perp}=C_9^\text{eff}+\frac{2m_b m_B}{q^2}C_7^\text{eff}, \qquad
C_{9,\parallel}=C_9^\text{eff}+\frac{2m_b}{m_B}C_7^\text{eff}.
\end{equation}
We denote the large-recoil expression 
as $P_5^\prime|_\infty$ and leave aside non-local terms, corresponding to non-factorisable corrections. Our result agrees with Eq.~(25) of Ref.~\cite{Jager:2014rwa} for the terms proportional to
$a_{V_-},a_{T_-},a_{V_0},a_{T_0}$, but we find an additional term proportional to $a_{V_+}$. 
We would like to stress that precisely this term, which is hidden in ``further terms'' and not discussed in 
Ref.~\cite{Jager:2014rwa}, dominates the power corrections in the anomalous region around $q_0^2\sim 6$\,GeV$^2$, as can be seen from the numerical evaluation of Eq.~(\ref{eq:analyticP5p}):
\begin{align}
P_5^\prime(6\,\text{GeV}^2)=P_5^\prime|_\infty(6\,\text{GeV}^2)\Bigg(1&+0.18\dfrac{2a_{V_-}-2a_{T_-}}{\xi_\perp}
-0.73\dfrac{2a_{V_+}}{\xi_\perp}+0.02\dfrac{2a_{V_0}-2a_{T_0}}{\tilde{\xi}_\parallel}\nonumber\\&+\text{nonlocal terms}\Bigg)+O\left(\dfrac{m_{K^*}}{m_B},\dfrac{\Lambda^2}{m_B^2},\dfrac{q^2}{m_B^2}\right).\label{eq:numericP5p}
\end{align}
This means that the discussion on the scheme dependence of $P_5^\prime$ in Ref.~\cite{Jager:2014rwa} only takes into account numerically subleading contributions.
Converted into the transversity basis, Eq.~(\ref{eq:analyticP5p}) becomes 
\begin{align}
P_5^\prime=P_5^\prime|_\infty\Bigg(1&+\dfrac{a_{A_1}+a_V-2a_{T_1}}{\xi_\perp}\dfrac{C_7^\text{eff}(C_{9,\perp}C_{9,\parallel}-C_{10}^2)}{(C_{9,\perp}+C_{9,\parallel})(C_{9,\perp}^2+C_{10}^2)}\dfrac{m_b m_B}{q^2}\nonumber\\
&-\dfrac{a_{A_1}-a_V}{\xi_\perp}\dfrac{C_{9,\parallel}}{C_{9,\perp}+C_{9,\parallel}}-\dfrac{a_{T_1}-a_{T_3}}{\tilde{\xi}_\parallel}
\dfrac{C_7^\text{eff}(C_{9,\perp}C_{9,\parallel}-C_{10}^2)}{(C_{9,\perp}+C_{9,\parallel})(C_{9,\parallel}^2+C_{10}^2)}\dfrac{m_b}{m_{K^*}}\\
&+\text{nonlocal terms}\Bigg)+O\left(\dfrac{m_{K^*}}{m_B},\dfrac{\Lambda^2}{m_B^2}\right),\nonumber
\label{eq:analyticP5ptrans}
\end{align}
with the dominant term being proportional to the combination $a_{A_1}-a_V$ of power correction parameters. 
If $A_1$ or $V$ is chosen as input for $\xi_\perp$, the corresponding parameter $a_{A_1}$ or $a_V$
vanishes identically. On the other hand, if $T_1$ is taken as input, both 
$a_{A_1}$ or $a_V$ survive and their independent variation leads to an increase of the errors associated to power corrections. This behaviour explains part of the inflated errors in Ref.~\cite{Jager:2012uw} and
it is analytically pinned down in Eqs.~(\ref{eq:analyticP5p}) and (\ref{eq:analyticP5ptrans}). The formulae support the numerical analysis reported in Fig.~2 of Ref.~\cite{Descotes-Genon:2014uoa}, where the binned predictions for $P_1,P_2,P_4',P_5'$ were given in the two schemes with $\xi_\perp$ defined from $V$ or $T_1$, respectively.
Without any further assumption on the correlations between the parameters 
$a_F$, Eqs.~(\ref{eq:analyticP5p}) and (\ref{eq:analyticP5ptrans}) manifest an explicit scheme dependence whose origin and interpretation was discussed in detail in Sec.~\ref{sec:schemedep}. 

For the observable $P_1$, which vanishes in the large-recoil limit, we find in the helicity basis
\begin{eqnarray}
P_1&=&-\dfrac{2a_{V_+}}{\xi_\perp}\,\dfrac{(C_9^\text{eff}C_{9,\perp}+C_{10}^2)}{C_{9,\perp}^2+C_{10}^2}-\dfrac{2b_{T_+}}{\xi_\perp}\,
\dfrac{2C_7^\text{eff}C_{9,\perp}}{C_{9,\perp}^2+C_{10}^2}\frac{m_b}{m_B}\nonumber\\
& &\qquad+\text{nonlocal terms}+O\left(\dfrac{m_{K^*}}{m_B},\dfrac{\Lambda^2}{m_B^2},\dfrac{q^2}{m_B^2}\right),
\label{eq:analyticP1}
\end{eqnarray}
turning in the transversity basis into
\begin{eqnarray}
P_1&=&-\dfrac{a_{A_1}-a_V}{\xi_\perp}\,\dfrac{(C_9^\text{eff}C_{9,\perp}+C_{10}^2)}{C_{9,\perp}^2+C_{10}^2}-\dfrac{b_{T_2}-b_{T_1}}{\xi_\perp}\,
\dfrac{2C_7^\text{eff}C_{9,\perp}}{C_{9,\perp}^2+C_{10}^2}\frac{m_b}{m_B}\nonumber\\
&&\qquad+\text{nonlocal terms}+O\left(\dfrac{m_{K^*}}{m_B},\dfrac{\Lambda^2}{m_B^2},\dfrac{q^2}{m_B^2}\right).
\label{eq:analyticP1bis}
\end{eqnarray}
Our result, Eq.~(\ref{eq:analyticP1}), fully agrees with Eq.~(26) of Ref.~\cite{Jager:2014rwa}. The authors of Ref.~\cite{Jager:2014rwa} used this result to argue that $P_1$ should be much cleaner than $P_5^\prime$ because it only involves one soft form factor and a lower number of power correction parameters $a_F$. However, the total number of power correction parameters is not the relevant criterion to decide whether an observable is clean:
as seen before, in the case of $P_5^\prime$ the coefficients in front of the power correction parameters exhibit a strong hierarchy, 
so that in practice only one term becomes relevant. As a matter of fact,  the leading power corrections for both $P_5^\prime$ and $P_1$ stem from 
$a_{V_+}$ and the respective coefficients are of the same size, as seen when comparing the evaluation of 
Eq.~(\ref{eq:analyticP1}) for $q^2_0=6$\,GeV$^2$,
\begin{align}
P_1(6\,\text{GeV}^2)=&-1.21\dfrac{2a_{V_+}}{\xi_\perp}+0.05\dfrac{2b_{T_+}}{\xi_\perp}\,+\text{nonlocal terms}+O\left(\dfrac{m_{K^*}}{m_B},\dfrac{\Lambda^2}{m_B^2},\dfrac{q^2}{m_B^2}\right),
\end{align}
with the corresponding one for $P_5^\prime$ from Eq.~(\ref{eq:numericP5p}). Therefore, $P_1$ and $P_5^\prime$
are on an equal footing with respect to power corrections, and all statements above, regarding the scheme 
dependence of $P_5^\prime$, also apply to $P_1$. Like $P_5^\prime$, $P_1$ suffers from an increase
of power corrections when $\xi_\perp$ is defined from $T_1$ instead of from $V$, as already demonstrated
numerically in Fig.~2 of Ref.~\cite{Descotes-Genon:2014uoa} and  analytically in Eq.(\ref{eq:analyticP1bis}).

Turning finally to the observable $P_2$, we find in the helicity basis
\begin{align}
P_2=P_2|_\infty\Bigg(1&+\dfrac{2a_{V_-}-2a_{T_-}}{\xi_\perp}\,\dfrac{C_7^\text{eff}(C_{9,\perp}^2-C_{10}^2)}{C_{9,\perp}(C_{9,\perp}^2+C_{10}^2)}\dfrac{m_b m_B}{q^2}+\text{nonlocal terms}\Bigg)\nonumber\\
&+O\left(\dfrac{m_{K^*}}{m_B},\dfrac{\Lambda^2}{m_B^2},\dfrac{q^2}{m_B^2}\right),
\end{align}
which translates into
\begin{align}
P_2=P_2|_\infty\Bigg(1&+\dfrac{a_V+a_{A_1}-a_{T_1}-a_{T_2}}{\xi_\perp}\,\dfrac{C_7^\text{eff}(C_{9,\perp}^2-C_{10}^2)}{C_{9,\perp}(C_{9,\perp}^2+C_{10}^2)}\dfrac{m_b m_B}{q^2}+\text{nonlocal terms}\Bigg)\nonumber\\
&+O\left(\dfrac{m_{K^*}}{m_B},\dfrac{\Lambda^2}{m_B^2},\dfrac{q^2}{m_B^2}\right),
\end{align}
in the transversity basis, with $P_2|_\infty=C_{9,\perp}C_{10}/(C_{9,\perp}^2+C_{10}^2)$. Unlike $P_1$ and $P_5^\prime$,
the leading term in $P_2$ involves both (axial-)vector and tensor power corrections, and at first sight it seems that
there is no preference whether to define $\xi_\perp$ from $V$ or from $T_1$. Note, however, that the 
kinematic relation $T_1(0)=T_2(0)$ implies $a_{T_1}=a_{T_2}$ and that a definition from $T_1$ hence absorbs
both $a_{T_1}$ and $a_{T_2}$ and leads to smaller uncertainties from corrections. Again, this is confirmed by the numerical results in Fig.~2 of Ref.~\cite{Descotes-Genon:2014uoa}.

We see that the scheme dependence of the angular observables can be explicitly worked out by studying the analytic dependence 
on the power correction parameters. Our results agree with Ref.~\cite{Jager:2014rwa} for $P_1$, 
but we have shown that the formula for $P_5^\prime$ in Ref.~\cite{Jager:2014rwa} actually misses the dominant 
and manifestly scheme-dependent term. Our analytic formulae allow us to understand how different schemes can yield significantly different uncertainties 
if one treats power corrections as uncorrelated, in perfect agreement with the numerical discussion in Ref.~\cite{Descotes-Genon:2014uoa}.
We can spot the relevant form factor(s) whose power corrections are going to have the main impact on each observable, 
and thus identify appropriate schemes to compute each observable accurately.

\section{Reassessing the reappraisal of long-distance charm loops} \label{sec:nonfacPC}

We now turn to the second main source of hadronic uncertainties: non-factorisable $\Lambda_{\textrm{QCD}}/m_B$ corrections associated with non-perturbative $c\bar{c}$ loops. 
Since these contributions can mimic a shift in the Wilson coefficient $C_9$, one may wonder
 how to disentangle them from possible short-distance new physics.
While the latter would induce a $q^2$-independent $C_9$, universal for the three different transversities $i=\perp,\|,0$, 
non-factorisable long-distance effects from $c\bar{c}$ loops in general introduce a $q^2$- and transversity dependence 
that can be cast into effective coefficient functions $C_{9 \, i}^{c\bar{c}}(q^2)$. A promising strategy thus consists in 
investigating whether the $B\to K^*\mu^+\mu^-$ data points towards a $q^2$-dependent effect. To this end the authors of Refs.~\cite{Ciuchini:2015qxb,Ciuchini:2016weo}
performed a fit of the functions $C_{9 \, i} ^{c\bar{c}}(q^2)$ to the data using a polynomial parametrisation. In Sec.~\ref{sec:Ciu} we comment on the results,
before presenting in Sec.~\ref{sec:ccfit} our own analysis based on a different, frequentist, statistical framework.

\subsection{A thorough interpretation of the results of Refs.~\cite{Ciuchini:2015qxb,Ciuchini:2016weo}}
\label{sec:Ciu}

The analysis in Refs.~\cite{Ciuchini:2015qxb,Ciuchini:2016weo} introduces for each helicity $\lambda=0,\pm 1$ 
a second-order polynomial in $q^2$:
\begin{equation} 
h_\lambda=h_\lambda^{(0)} + \frac{q^2 }{1\ {\rm GeV}^2} h_\lambda^{(1)} + \frac{q^4}{1\ {\rm GeV}^4}h_\lambda^{(2)} \label{hlambda}.
\end{equation}
The functions $h_\lambda$, with a total number of 18 real parameters, then enter the $B\to K^*\mu^+\mu^-$ transversity amplitudes as follows:
\begin{eqnarray}
A^0_{L,R}&=& A^0_{L,R}(s_i=0) + \frac{N}{q^2} \left(
 \frac{q^2 }{1\ {\rm GeV}^2} h_0^{(1)} + \frac{q^4}{1\ {\rm GeV}^4}h_0^{(2)} \right),\nonumber\\
A^{\|}_{L,R}&=& A^{\|}_{L,R}(s_i=0) \nonumber\\
&&\quad+ \frac{N}{\sqrt{2}q^2}\left[ (h_+^{(0)}+h_-^{(0)}) + \frac{q^2 }{1\ {\rm GeV}^2} (h_+^{(1)}+h_-^{(1)}) + \frac{q^4}{1\ {\rm GeV}^4}(h_+^{(2)}+h_-^{(2)})\right],\nonumber \\
A^{\perp}_{L,R}&=& A^{\perp}_{L,R}(s_i=0) \nonumber\\
&&\quad+ \frac{N}{\sqrt{2}q^2}\left[ (h_+^{(0)}-h_-^{(0)}) + \frac{q^2 }{1\ {\rm GeV}^2} (h_+^{(1)}-h_-^{(1)}) + \frac{q^4}{1\ {\rm GeV}^4}(h_+^{(2)}-h_-^{(2)})\right],\label{hpara} 
\end{eqnarray}
with the normalisation
\begin{equation}
N=V_{tb}V_{ts}^* \frac{m_B^{3/2}G_F\alpha\sqrt{q^2}}{\sqrt{3\pi}}  \lambda^{1/4}(m_B^2,m_{K^*}^2,q^2) \left(1-\frac{4m_\ell^2}{q^2}\right)^{1/4}.
\end{equation}
Here, $s_i=0$ indicates that only the perturbative quark-loop contribution $Y(q^2)$ has been included in the amplitudes $A^\lambda_{L,R}(s_i=0)$
while any long-distance contribution as the one calculated in Ref.~\cite{Khodjamirian:2010vf} and included in Ref.~\cite{Descotes-Genon:2015uva} is switched off.

The coefficients $h_\lambda^{(i)}$ parametrise the $q^2$-expansion of the charm-loop contribution to the various helicity amplitudes, but can also (partially) be mimicked by NP contributions to the Wilson coefficients $C_7$ and $C_9$.
Note that a NP contribution to $C_7$ would yield a pole at $s=0$ and thus contribute to $h^{(0)}_\lambda$ and higher orders, whereas a NP contribution to $C_9$ would contribute only starting from $h^{(1)}_\lambda$ and higher orders. Let us stress that both kinds of NP contributions would also contribute to $h^{(2)}_\lambda$, since they enter the transversity amplitudes as a Wilson coefficient multiplied by a $q^2$-dependent form factor~\footnote{It is thus not correct to state that $h^{(2)}$ and higher coefficients can arise only due to long-distance physics as suggested in Ref.~\cite{Ciuchini:2015qxb,Ciuchini:2016weo}. Even though the form factors do not vary strongly with $q^2$, the presence of NP contributions to Wilson coefficients would generate terms corresponding to (small) contributions to higher orders in the polynomial expansion.}. Contrary to Refs.~\cite{Ciuchini:2015qxb,Ciuchini:2016weo}, we have set $h_0^{(0)}=0$ in order to avoid an unphysical pole at $q^2=0$ in $A^0_{L,R}$ (which for instance would result in a divergence in $BR(B\to K^*\gamma)$).

For a proper interpretation of the results obtained in Ref.~\cite{Ciuchini:2015qxb}, it is important to note that the authors study two different hypotheses:
\begin{itemize}
\item Hypothesis 1: No constraint is imposed on the long-distance charm-loop contribution represented by the coefficients $h_\lambda^{(i)}$, and the 
results of the LCSR computation in Ref.~\cite{Khodjamirian:2010vf} are not used in the fit. 
Instead, after fitting the functions $h_\lambda(q^2)$ to the $B\to K^*\mu^+\mu^-$ data they are compared with the functions
${\tilde g}_i^{\cal M}$ calculated in Ref.~\cite{Khodjamirian:2010vf}.
We have checked  the relation between the functions $\tilde{g}_i^\mathcal{M}$ and the long-distance charm-loop contributions $h_\lambda$, given by Eq.~(2.7) in Ref.~\cite{Ciuchini:2015qxb} (up to the correction $C_1 \to C_2$ noticed in Ref.~\cite{Ciuchini:2016weo}). 
Rewriting  the amplitudes $\mathcal{M}_{1,2,3}$ in Ref.~\cite{Khodjamirian:2010vf} in terms of helicity amplitudes leads to~\footnote{Even though Eq.~(\ref{translation}) is also valid for the imaginary part of the functions, we only consider the real part of the ${\tilde g}_i^{\cal M}$ here, as 
the authors of Ref.~\cite{Khodjamirian:2010vf} consider these contributions to be real in the region of interest within their approximations.}:
\begin{eqnarray} \label{translation}
{\rm Re} \, {\tilde g}^{\cal M}_1&=& -\frac{1}{2 C_2} \frac{16 m_B^3 (m_B+ m_{K^*}) \pi^2}{\sqrt{\lambda(q^2)} V(q^2) q^2} \left({\rm Re}\, h_{-}(q^2) - {\rm Re}\, h_{+}(q^2) \right), \nonumber\\
{\rm Re} \, {\tilde g}^{\cal M}_2&=& -\frac{1}{2 C_2} \frac{16 m_B^3  \pi^2}{(m_B+ m_{K^*}) A_1(q^2) q^2} \left({\rm Re}\, h_{-}(q^2) + {\rm Re}\, h_{+}(q^2) \right), \nonumber\\
{\rm Re} \, {\tilde g}^{\cal M}_3&=& \frac{1}{2 C_2} \frac{64 \pi^2 m_B^3  m_{K^*}\sqrt{q^2} (m_B+ m_{K^*})  }{\lambda (q^2) A_2(q^2) q^2} [{\rm Re}\, h_{0}(q^2)  
\nonumber\\  &-& \frac{16 m_B^3 \pi^2 (m_B+m_{K^*}) (m_B^2 -q^2 -m_{K^*}^2)}{\lambda (q^2) A_2(q^2) q^2} \left( {\rm Re}\, h_{-}(q^2) + {\rm Re}\, h_{+}(q^2) \right) ].
\end{eqnarray}
 It is interesting to observe that the results of the fit in Ref.~\cite{Ciuchini:2015qxb} for ${\tilde g}^{\cal M}_i$ seem to agree well with the LCSR estimates of Ref.~\cite{Khodjamirian:2010vf} if in all amplitudes approximately the same $q^2$-independent shift is added to the LCSR result. This observation is in line with the conclusions from global fits~\cite{Altmannshofer:2014rta,Altmannshofer:2015sma,Descotes-Genon:2015uva}, bearing in mind that in Ref.~\cite{Ciuchini:2015qxb,Ciuchini:2016weo} basically only $B\to K^*\mu^+\mu^-$ data is used and that the authors interpret this constant shift as being of hadronic origin. Notice that such a $q^2$-independent shift (very similar for all helicity amplitudes) is at odds with a $q^2$- and helicity-dependent contribution expected in the case of an hadronic effect, in particular if it is attributed to tails of resonances.
Note, however, that a firm conclusion can only be drawn by comparing the quality of a fit for a $q^2$-independent contribution with the one for $q^2$-dependent functions, a task that was not carried out in Refs.~\cite{Ciuchini:2015qxb,Ciuchini:2016weo} and that will be performed in Sec.~\ref{sec:ccfit}.
In any case, one should keep in mind that a universal shift in $C_9^\mu$ due to NP can also explain the deviations in $B_s\to \phi\mu^+\mu^-$ and the violation of lepton-flavour universality suggested by $R_K$ and $Q_5={P_5^{\mu}}'-{P_5^{e}}'$, which is not the case for hadronic $c\bar{c}$ contributions.
 
 \item Hypothesis 2: In a second analysis, the authors of Ref.~\cite{Ciuchini:2015qxb} impose an additional constraint to the fit: they assume that the results of Ref.~\cite{Khodjamirian:2010vf} 
 hold exactly for $q^2 \leq 1$ GeV$^2$, while they do not make any assumptions for $q^2>1$ GeV$^2$ and again set all the Wilson coefficients to their SM value. 
 The results obtained in this second approach have to be interpreted with great care:
 \begin{itemize}
 \item[i)] The authors of Ref.~\cite{Ciuchini:2015qxb}  decide to take the results of Ref.~\cite{Khodjamirian:2010vf} as exact in the region $q^2<1$ GeV$^2$ but to discard them for larger $q^2$: this choice of range is rather arbitrary, as the LCSR approach yields a computation valid up to 2 GeV$^2$ according to Ref.~\cite{Khodjamirian:2010vf}, and the extrapolation via the dispersion relation is deemed appropriate up to 4 GeV$^2$ by the authors of Ref.~\cite{Ciuchini:2015qxb} themselves.
 \item[ii)] The additional constraint artificially tilts the fit by forcing it to follow a behaviour at $q^2 \lesssim  1$ GeV$^2$ against the trend of data (which would prefer to have a constant shift $C_9^{\rm NP}$, as discussed in Refs.~\cite{Descotes-Genon:2015uva,Altmannshofer:2014rta,Altmannshofer:2015sma,Hurth:2014vma}, corresponding to non-vanishing $h^{(1)}_\lambda$ in the framework of Ref.~\cite{Ciuchini:2015qxb}). It is compensated by a spurious $q^4$-dependence with $h^{(2)}_{\lambda}\neq 0$, which is then interpreted in Refs.~\cite{Ciuchini:2015qxb,Ciuchini:2016weo} as an indication of non-local hadronic effects.
\item[iii)]  In the region below 1 GeV$^2$, the treatment of the distribution by LHCb means that the data correspond to slightly different observables from the optimized observables defined in Ref. \cite{Matias:2012xw, Descotes-Genon:2013vna}, as discussed in Sec. 2.3.1 in Ref.~\cite{Descotes-Genon:2015uva}  and below. This effect, which can be taken into account by a redefinition of the optimized observables, is not considered in Ref.~\cite{Ciuchini:2015qxb,Ciuchini:2016weo}  and can affect the outcome of the analysis.
 \item[iv)] Finally, the LCSR computation of Ref.~\cite{Khodjamirian:2010vf} does not take into account all non-local effects but is an estimate of the soft gluon part with respect to the leading-order factorisable contribution, from which the imaginary part is still missing. In this sense it is not consistent to compare the absolute value of the fitted  ${\tilde g}_i^{\cal M}$ obtained from data with the computation of Ref.~\cite{Khodjamirian:2010vf}, and if one still insists in doing so (ignoring all previous issues), at least one should compare their real parts rather than the absolute values.
\end{itemize}
\end{itemize}
We conclude that a fit under the second hypothesis cannot indicate whether a $q^2$-dependent effect is favoured over a constant one, since 
it artificially creates a $q^2$-dependence by putting a constraint on one side (below $q^2=1$ GeV$^2$). A fit under the first hypothesis 
can be an appropriate method, but requires to compare the quality of the fits obtained in both cases under consideration of the number of free parameters.
We will address this issue in the following.

\subsection{A frequentist fit}
\label{sec:ccfit}

We are going to perform fits using the approach described in Ref.~\cite{Descotes-Genon:2015uva}, taking LHCb data on $B\to K^*\mu\mu$ as data. 
We follow the theoretical framework of Ref.~\cite{Descotes-Genon:2015uva} for the predictions of the observables, but 
modify it slightly to remain as close as possible to the fits shown in Refs.~\cite{Ciuchini:2015qxb,Ciuchini:2016weo}: we will not use
 the computation of long-distance charm effects in Ref.~\cite{Khodjamirian:2010vf}. In practice, this amounts to keeping only the perturbative function $Y(q^2)$ while setting all three $s_i=0$. We treat the form factors using the soft-form-factor approach with the inputs of Ref.~\cite{Khodjamirian:2010vf}, and employ the same parametrisation Eq.~(\ref{hpara}) as Refs.~\cite{Ciuchini:2015qxb,Ciuchini:2016weo} for the long-distance charm contribution, extending it in a straightforward way to the order $q^6$ by introducing the parameters $h^{(3)}_\lambda$. We take all coefficients of the expansion as real, following Ref.~\cite{Khodjamirian:2010vf}. Note that the results of Ref.~\cite{Ciuchini:2015qxb,Ciuchini:2016weo} favour mostly real values for $h_+$ and $h_0$, but not necessarily for $h_-$.

Our fits differ from the ones in Refs.~\cite{Ciuchini:2015qxb,Ciuchini:2016weo} with respect to the statistical framework.
We use a frequentist approach and in particular do not assume any a-priori range for the fit parameters $h_\lambda^{(i)}$, contrary to the Bayesian approach in Refs.~\cite{Ciuchini:2015qxb,Ciuchini:2016weo} where (flat or Gaussian) priors are used for the polynomial parameters.
Keeping in mind that the functions $h_\lambda(q^2)$ are expansions in $q^2$, we perform fits 
allowing for $h_\lambda^{(i)}$ with $i\leq n$, increasing progressively the degree of the polynomials $n$. At each order, we determine the minimum $\chi^2_{\rm min}$ as well as the difference between the $\chi^2_{\rm min}$ with polynomial degrees $n-1$ and $n$, and the pull of the hypothesis $h_{0,+,-}^{(n)}=0$. This information indicates 
the improvement of the fit obtained by increasing the degree of the polynomial expansion.

In Tabs.~\ref{tab:A-SM} and \ref{tab:A-C9}, we provide the results in the SM case and in the NP scenario $C_9^{\textrm{NP}}=-1.1$, respectively, using only $B\to K^*\mu^+\mu^-$ data. We see that in both cases, the fit clearly improves when increasing the degree of the polynomial from $n=0$ to $n=1$ (the addition of the parameters $h^{(1)}_\lambda$
leads to a $q^2$ dependence similar to that of a NP contribution to the Wilson coefficient $C_9$). On the other hand, including quadratic or cubic terms does not provide any significant improvement. This implies that the fit does not hint at a $q^2$-dependence beyond the one generated by the Wilson coefficients $C_7$ and $C_9$. 
\begin{table}[t]
\begin{center}
{\footnotesize\begin{tabular}{c|cc|cc|cc|cc}  $n$&\multicolumn{2}{c|}{0}&\multicolumn{2}{c|}{1}&\multicolumn{2}{c|}{2}&\multicolumn{2}{c}{3}\\ 
\hline
$\chi^{2(n)}_{\rm min}$ & \multicolumn{2}{c|}{70.00} & \multicolumn{2}{c|}{52.70} & \multicolumn{2}{c|}{51.50} & \multicolumn{2}{c}{51.20} \\ 
$\chi^{2(n-1)}_{\rm min}-\chi^{2(n)}_{\rm min}$  & 1.64 &  (0.5 $\sigma$) & 17.30 & (3.4 $\sigma$) & 1.14 & (0.3 $\sigma$) & 0.35 & (0.1 $\sigma$)\\ 
$h_+^{(0)}$ & $0.17^{+1.15}_{-0.62}$ & (0.3 $\sigma$) & $2.22^{+1.07}_{-1.13}$ & (2.0 $\sigma$) & $1.28^{+1.45}_{-0.40}$ & (3.2 $\sigma$) & $1.19^{+1.32}_{-0.62}$ & (1.9 $\sigma$)\\ 
$h_+^{(1)}$ & &  & $-2.37^{+1.42}_{-0.57}$ & (1.7 $\sigma$) & $-1.66^{+1.43}_{-1.03}$ & (1.2 $\sigma$) & $-1.31^{+0.83}_{-1.21}$ & (1.6 $\sigma$)\\ 
$h_+^{(2)}$ & &  & &  & $-0.11^{+0.19}_{-0.14}$ & (0.6 $\sigma$) & $-0.09^{+0.11}_{-0.11}$ & (0.8 $\sigma$)\\ 
$h_+^{(3)}$ & &  & &  & &  & $-0.00^{+0.01}_{-0.00}$ & (0.2 $\sigma$)\\ 
$h_-^{(0)}$ & $1.30^{+1.47}_{-1.07}$ & (1.2 $\sigma$) & $2.62^{+1.58}_{-2.69}$ & (1.0 $\sigma$) & $2.30^{+1.68}_{-1.76}$ & (1.3 $\sigma$) & $1.85^{+1.93}_{-1.09}$ & (1.7 $\sigma$)\\ 
$h_-^{(1)}$ & &  & $-0.34^{+0.90}_{-0.53}$ & (0.4 $\sigma$) & $-1.24^{+1.53}_{-0.21}$ & (0.8 $\sigma$) & $-0.94^{+1.19}_{-0.64}$ & (0.8 $\sigma$)\\ 
$h_-^{(2)}$ & &  & &  & $0.13^{+0.06}_{-0.19}$ & (0.7 $\sigma$) & $0.11^{+0.12}_{-0.18}$ & (0.6 $\sigma$)\\ 
$h_-^{(3)}$ & &  & &  & &  & $0.00^{+0.00}_{-0.01}$ & (0.0 $\sigma$)\\ 
$h_0^{(1)}$ & &  & $-1.00^{+1.69}_{-0.89}$ & (0.6 $\sigma$) & $-1.35^{+1.70}_{-1.14}$ & (0.8 $\sigma$) & $-0.96^{+1.01}_{-1.45}$ & (0.9 $\sigma$)\\ 
$h_0^{(2)}$ & &  & &  & $0.10^{+0.12}_{-0.10}$ & (1.0 $\sigma$) & $0.11^{+0.11}_{-0.17}$ & (0.6 $\sigma$)\\ 
$h_0^{(3)}$ & &  & &  & &  & $-0.00^{+0.01}_{-0.00}$ & (0.2 $\sigma$)\\ 
\end{tabular}}
\end{center}
\caption{Fit to $B\to K^*\mu^+\mu^-$ only, with $C_9^{\mu, \rm NP}=0$, using LCSR from Ref.~\cite{Khodjamirian:2010vf} in the soft-form-factor approach 
employed by Ref.~\cite{Descotes-Genon:2015uva}. All coefficients are given in units of $10^{-4}$. Different orders $n$ of the polynomial parametrisation of the long-distance charm-loop contribution are considered. If this contribution is set to zero, the fit yields $\chi^2_{{\rm min;}h=0}=71.60$  for $N_{dof}=59$.}\label{tab:A-SM}
\end{table}
In Refs.~\cite{Ciuchini:2015qxb,Ciuchini:2016weo} a different $q^2$-dependence was advocated referring to the parameter $h_-^{(2)}$ which showed a $\lesssim 2\sigma$
deviation from $h_-^{(2)}=0$.  We would like to emphasize that it is impossible to draw conclusions from a single parameter and that a global assessment of the whole fit is required. For instance, from our tables one can see that 
increasing the order of the expansion can lead to a reshuffling of the overall deviation from zero of the functions $h_\lambda(q^2)$ among the various 
expansion parameters, even in the case that no significant improvement of the fit is obtained. For instance, in the SM fit (Tab.~\ref{tab:A-SM}) 
the parameter $h^{(0)}_+$ deviates from zero by $1.3\sigma$ at the order $n=2$, but by $2.8\sigma$ at $n=3$. We would expect a similar analysis to be possible in the Bayesian framework proposed in Ref.~\cite{Ciuchini:2015qxb}, by comparing the information criteria for
the two hypotheses ``no constraint for $q^2\leq 1$ GeV$^2$ and  $h^{(2)}_\lambda$ left free" and  ``no constraint for $q^2\leq 1$ GeV$^2$ and $h^{(2)}_\lambda=0$", which is unfortunately not provided in Ref.~\cite{Ciuchini:2015qxb}.

In the SM fit we find the pattern 
\begin{equation}
 h_+^{(0)}\geq 0,\quad\quad h_-^{(0)}\geq 0,\quad\quad
h_0^{(1)}\simeq 0,\quad\quad h_+^{(1)}\leq 0,\quad\quad h_-^{(1)}\simeq 0,
\end{equation}
while higher orders are compatible with zero. These findings are in rough agreement with Refs.~\cite{Ciuchini:2015qxb,Ciuchini:2016weo} 
for the $\lambda=0,+$ helicities. The differences can be attributed to the different treatment and input for the form factors and to the differences in the statistical approach. The comparison cannot be done easily for the $\lambda=-$ helicity, as large phases were found in Ref.~\cite{Ciuchini:2015qxb} whereas we considered only real $c\bar{c}$ contributions.

Setting $C_9^{\mu,{\rm NP}}=-1.1$ improves the $\chi^2_{\rm min}$ significantly without modifying the above conclusions (see Tab.~\ref{tab:A-C9}). As mentioned before, it is not strictly equivalent to modify $h^{(1)}$ or $C_9$ since the latter is multiplied by a $q^2$-dependent form factor. Therefore the results of the fits are not exactly identical, both for the $\chi^2_{\rm min}$ and the values of the expansion coefficients $h^{(n)}$ (this explains why the addition of $h^{(1)}$ still brings some improvement to the fit with $C_9^{\mu,{\rm NP}}=-1.1$, although more modestly than in the SM case).
In Tab.~\ref{tab:C}, we present the same fit as in Tab.~\ref{tab:A-SM} ($B\to K^*\mu^+\mu^-$ only, no NP contributions to the Wilson coefficients), taking the LCSR results from Ref.~\cite{Straub:2015ica} within the full-form factor approach. As can be seen from the comparison of the two tables, the same conclusions hold independently of the specific input for the form factors.

We also performed another fit (Tab.~\ref{tab:B}) where we consider the SM case but include all the exclusive $b\to se^+e^-$ and $b\to s\mu^+\mu^-$ observables discussed in Ref.~\cite{Descotes-Genon:2015uva}. We take the same parameters for the charm-loop contributions in $B_s\to \phi\ell^+\ell^-$ and 
$B\to K^*\ell^+\ell^-$ (i.e., we assume an $SU(3)$ flavour symmetry for this long-distance contribution), but we neglect the effect of charm loops in $B\to K\ell^+\ell^-$ (in agreement with Ref.~\cite{Khodjamirian:2010vf}). Compared to Ref.~\cite{Descotes-Genon:2015uva} and due to its direct relation with 
the charm-loop contribution, we have also added the $B\to K^*\gamma$ branching ratio that was not included in our earlier analyses (we have checked that including this observable does affect neither the outcome of the global fits presented in 
Ref.~\cite{Descotes-Genon:2015uva}, nor the fits presented in this section).
We see again that there is no strong for quadratic $h$ terms: $h_-^{(2)}$ prefers to be slightly different from zero (positive), but the data can also be
described equivalently well using only constant and linear contributions.

At this stage, we see that the data require constant and linear contributions, as expected also from Ref.~\cite{Khodjamirian:2010vf}. On the other hand, the data do not require additional 
quadratic or cubic contributions, contrary to the claim made in Ref.~\cite{Ciuchini:2015qxb}. This claim was later amended in Ref.~\cite{Ciuchini:2016weo}, indicating that a solution with 
$h^{(2)}=0$ also leads to acceptable Bayesian fits. Our own fits indicate that the current data do not show signs of a large and unaccounted for hadronic contribution from charm loops.

\begin{table}
\begin{center}
{\footnotesize\begin{tabular}{c|cc|cc|cc|cc}  $n$&\multicolumn{2}{c|}{0}&\multicolumn{2}{c|}{1}&\multicolumn{2}{c|}{2}&\multicolumn{2}{c}{3}\\ 
\hline
$\chi^{2(n)}_{\rm min}$ & \multicolumn{2}{c|}{62.10} & \multicolumn{2}{c|}{51.60} & \multicolumn{2}{c|}{50.50} & \multicolumn{2}{c}{50.00} \\ 
$\chi^{2(n-1)}_{\rm min}-\chi^{2(n)}_{\rm min}$  & 1.23 &  (0.3 $\sigma$) & 10.50 & (2.4 $\sigma$) & 1.14 & (0.3 $\sigma$) & 0.53 & (0.1 $\sigma$)\\ 
$h_+^{(0)}$ & $0.66^{+1.06}_{-0.55}$ & (1.2 $\sigma$) & $1.97^{+1.32}_{-0.51}$ & (3.8 $\sigma$) & $1.62^{+1.22}_{-1.00}$ & (1.6 $\sigma$) & $1.43^{+1.13}_{-0.80}$ & (1.8 $\sigma$)\\ 
$h_+^{(1)}$ & &  & $-1.92^{+0.81}_{-0.76}$ & (2.4 $\sigma$) & $-1.29^{+1.70}_{-1.75}$ & (0.8 $\sigma$) & $-1.45^{+1.40}_{-0.74}$ & (1.0 $\sigma$)\\ 
$h_+^{(2)}$ & &  & &  & $-0.16^{+0.24}_{-0.09}$ & (0.7 $\sigma$) & $-0.09^{+0.08}_{-0.16}$ & (1.2 $\sigma$)\\ 
$h_+^{(3)}$ & &  & &  & &  & $0.00^{+0.01}_{-0.00}$ & (0.0 $\sigma$)\\ 
$h_-^{(0)}$ & $-0.14^{+1.43}_{-0.93}$ & (0.1 $\sigma$) & $1.90^{+1.99}_{-1.64}$ & (1.2 $\sigma$) & $1.87^{+2.71}_{-1.31}$ & (1.4 $\sigma$) & $1.93^{+1.93}_{-0.93}$ & (2.1 $\sigma$)\\ 
$h_-^{(1)}$ & &  & $-0.81^{+0.68}_{-0.43}$ & (1.2 $\sigma$) & $-0.56^{+0.48}_{-1.32}$ & (1.2 $\sigma$) & $-0.65^{+0.59}_{-0.82}$ & (1.1 $\sigma$)\\ 
$h_-^{(2)}$ & &  & &  & $-0.04^{+0.22}_{-0.07}$ & (0.2 $\sigma$) & $-0.02^{+0.14}_{-0.10}$ & (0.1 $\sigma$)\\ 
$h_-^{(3)}$ & &  & &  & &  & $-0.00^{+0.00}_{-0.00}$ & (0.2 $\sigma$)\\ 
$h_0^{(1)}$ & &  & $-1.28^{+1.17}_{-1.24}$ & (1.1 $\sigma$) & $-2.24^{+1.64}_{-1.43}$ & (1.4 $\sigma$) & $-2.08^{+0.90}_{-1.38}$ & (2.3 $\sigma$)\\ 
$h_0^{(2)}$ & &  & &  & $0.08^{+0.17}_{-0.07}$ & (1.1 $\sigma$) & $0.16^{+0.17}_{-0.12}$ & (1.3 $\sigma$)\\ 
$h_0^{(3)}$ & &  & &  & &  & $-0.00^{+0.01}_{-0.00}$ & (0.5 $\sigma$)\\ 
\end{tabular}}
\end{center}
\caption{Fit to $B\to K^*\mu^+\mu^-$ only, with $C_9^{\mu,\rm NP}=-1.1$, using LCSR from Ref.~\cite{Khodjamirian:2010vf} in the soft-form-factor approach 
employed by Ref.~\cite{Descotes-Genon:2015uva}. All coefficients are given in units of $10^{-4}$. Different orders $n$ of the polynomial parametrisation of the long-distance charm-loop contribution are considered. If this contribution is set to zero, the fit yields
$\chi^2_{{\rm min;}h=0}=63.30$  for $N_{dof}=59$.}\label{tab:A-C9}
\end{table}

\begin{table}
\begin{center}
{\footnotesize\begin{tabular}{c|cc|cc|cccc} $n$&\multicolumn{2}{c|}{0}&\multicolumn{2}{c|}{1}&\multicolumn{2}{c}{2}\\
\hline
$\chi^{2(n)}_{\rm min}$ & \multicolumn{2}{c|}{65.50} & \multicolumn{2}{c|}{52.70} & \multicolumn{2}{c}{52.40} \\ 
$\chi^{2(n-1)}_{\rm min}-\chi^{2(n)}_{\rm min}$  & 4.31 &  (1.2 $\sigma$) & 12.80 & (2.8 $\sigma$) & 0.26 & (0.0 $\sigma$)\\ 
$h_+^{(0)}$ & $0.05^{+1.21}_{-0.71}$ & (0.1 $\sigma$) & $1.40^{+1.12}_{-0.69}$ & (2.0 $\sigma$) & $1.10^{+1.66}_{-0.40}$ & (2.8 $\sigma$)\\ 
$h_+^{(1)}$ & &  & $-0.82^{+0.76}_{-0.41}$ & (1.1 $\sigma$) & $0.09^{+0.49}_{-1.22}$ & (0.1 $\sigma$)\\ 
$h_+^{(2)}$ & &  & &  & $-0.16^{+0.32}_{-0.06}$ & (0.5 $\sigma$)\\ 
$h_-^{(0)}$ & $1.24^{+1.04}_{-0.55}$ & (2.2 $\sigma$) & $0.53^{+1.00}_{-0.75}$ & (0.7 $\sigma$) & $0.78^{+0.80}_{-0.60}$ & (1.3 $\sigma$)\\ 
$h_-^{(1)}$ & &  & $0.43^{+0.46}_{-0.26}$ & (1.6 $\sigma$) & $0.19^{+0.66}_{-0.78}$ & (0.2 $\sigma$)\\ 
$h_-^{(2)}$ & &  & &  & $0.04^{+0.16}_{-0.07}$ & (0.6 $\sigma$)\\ 
$h_0^{(1)}$ & &  & $0.31^{+1.03}_{-0.43}$ & (0.7 $\sigma$) & $0.66^{+1.97}_{-0.60}$ & (1.1 $\sigma$)\\ 
$h_0^{(2)}$ & &  & &  & $-0.07^{+0.15}_{-0.10}$ & (0.5 $\sigma$)\\ 
\end{tabular}}
\end{center}
\caption{Fit to $B\to K^*\mu^+\mu^-$ only, with $C_9^{\mu,{\rm NP}}=0$, using LCSR results from Ref.~\cite{Straub:2015ica} in the full-form-factor approach. All coefficients are given in units of $10^{-4}$. Different orders $n$ of the polynomial parametrisation of the long-distance charm-loop contribution are considered. If this contribution is set to zero, the fit yields $\chi^2_{{\rm min;}h=0}=69.80$  for $N_{dof}=59$.}\label{tab:C}
\end{table}

\begin{table}
\begin{center}
{\footnotesize\begin{tabular}{c|cc|cc|cccc} $n$&\multicolumn{2}{c|}{0}&\multicolumn{2}{c|}{1}&\multicolumn{2}{c}{2}\\
\hline
$\chi^{2(n)}_{\rm min}$ & \multicolumn{2}{c|}{96.50} & \multicolumn{2}{c|}{75.50} & \multicolumn{2}{c}{75.50} \\ 
$\chi^{2(n-1)}_{\rm min}-\chi^{2(n)}_{\rm min}$  & 1.53 &  (0.4 $\sigma$) & 20.90 & (3.9 $\sigma$) & 0.10 & (0.0 $\sigma$)\\ 
$h_+^{(0)}$ & $0.39^{+1.00}_{-0.52}$ & (0.7 $\sigma$) & $1.19^{+1.29}_{-0.42}$ & (2.8 $\sigma$) & $1.16^{+1.04}_{-0.27}$ & (4.3 $\sigma$)\\ 
$h_+^{(1)}$ & &  & $-0.45^{+0.66}_{-0.48}$ & (0.7 $\sigma$) & $-0.29^{+0.83}_{-0.94}$ & (0.4 $\sigma$)\\ 
$h_+^{(2)}$ & &  & &  & $0.02^{+0.17}_{-0.17}$ & (0.1 $\sigma$)\\ 
$h_-^{(0)}$ & $0.72^{+1.12}_{-0.67}$ & (1.1 $\sigma$) & $-0.21^{+1.05}_{-0.37}$ & (0.2 $\sigma$) & $0.19^{+0.87}_{-0.60}$ & (0.3 $\sigma$)\\ 
$h_-^{(1)}$ & &  & $0.29^{+0.53}_{-0.17}$ & (1.7 $\sigma$) & $-0.58^{+1.18}_{-0.17}$ & (0.5 $\sigma$)\\ 
$h_-^{(2)}$ & &  & &  & $0.12^{+0.06}_{-0.13}$ & (1.0 $\sigma$)\\ 
$h_0^{(1)}$ & &  & $1.54^{+0.75}_{-0.48}$ & (3.2 $\sigma$) & $1.66^{+0.50}_{-1.08}$ & (1.5 $\sigma$)\\ 
$h_0^{(2)}$ & &  & &  & $0.01^{+0.13}_{-0.08}$ & (0.1 $\sigma$)\\ 
\end{tabular}}
\end{center}
\caption{Fit to exclusive $b\to se^+e^-$ and $b\to s\mu^+\mu^-$ observables with $C_9^{\mu, {\rm NP}}=0$, using the same approach as in Ref.~\cite{Descotes-Genon:2015uva}. All coefficients are given in units of $10^{-4}$. Different orders $n$ of the polynomial parametrisation of the long-distance charm-loop contribution for $B\to V\ell^+\ell^-$ are considered. If this contribution is set to zero, the fit yields $\chi^2_{{\rm min;}h=0}=98.00$  for $N_{dof}=81$.} \label{tab:B}
\end{table}

\section{Further experimental tests of the role of hadronic uncertainties}\label{sec:tests} \label{sec:kinem}

A different approach to hadronic uncertainties consists in identifying observables and kinematic regions totally (or partially) free from some of these uncertainties.
Contributions from $c\bar{c}$ loops enter many $B\to K^*\ell\ell$ observables, but it is worth noticing that not all of them exhibit the same sensitivity to these effects. 

Let us start by recalling a few facts concerning the structure of this contribution.
The long-distance $c\bar{c}$ contribution has a $1/q^2$ pole due to the photon propagator:
following Ref.~\cite{Khodjamirian:2010vf}, we have absorbed this singular contribution into an effective $C_9$.
If only regular expressions (no poles) are preferred, one can split the $c\bar{c}$ contribution into two parts: the pole term affects $C_7$ and the remaining regular part enters $C_9$.

The Wilson coefficient $C_7$ (SM and NP) is accurately extracted from the inclusive branching ratio $BR(B \to X_s \gamma)$, where hadronic effects are tightly controlled, providing a slight preference for a narrow negative  range for $C_7^{\rm NP}$ if only NP is allowed in this coefficient (see Refs.~\cite{Misiak:2015xwa} and \cite{Descotes-Genon:2015uva}). The comparison between this inclusive observable and exclusive observables that contain long-distance charm contributions (like $BR(B^+ \to K^{*+} \gamma)$ and $BR(B^0 \to K^{*0} \gamma)$) does not leave much space for a sizeable long-distance charm contribution at $q^2=0$ entering $C_7$. The sum of the NP and long-distance charm contributions favours a negative contribution, increasing in absolute value the size of $C_7^{\rm SM}=-0.29$ (see for instance Ref.~\cite{Paul:2016urs}). 
Indeed the allowed ranges for $C_7$ and $C_7^\prime$  found in Ref.~\cite{Paul:2016urs} (see Fig. 2) are in very good agreement with the results of the global fit shown in Fig.10 of Ref.~\cite{Descotes-Genon:2015uva} under similar conditions (keeping $C_9^{\rm NP}=0$). 

We can also illustrate this expectation of very small contributions by considering the charm-loop parametrisation introduced in Sec.~\ref{sec:nonfacPC}. The long-distance charm  contribution to $C_7$ for the transverse amplitudes can be expressed as \cite{Ciuchini:2015qxb}
\begin{eqnarray}\label{c7charm}
C_{7 \, \perp}^{c\bar{c}}&=& \frac{8 \pi^2 m_B^3}{\sqrt{\lambda(0)}
m_b T_1(0)} \left( h_{+}^{(0)}-h_{-}^{(0)} \right),  \cr
C_{7 \, \|}^{c\bar{c}}&=& -\frac{8 \pi^2 m_B^3}{\sqrt{\lambda(0)}
m_b T_1(0)} \left( h_{+}^{(0)}+h_{-}^{(0)} \right). 
\end{eqnarray}
 Using the values for the charm contribution obtained from the fit  from Tab.~\ref{tab:A-SM} (SM) and Tab.~\ref{tab:A-C9} ($C_9^{\rm NP}=-1.1$) in the optimal case $n=1$ one can determine these contributions (see
 Table \ref{tab:c7cc})~\footnote{
  Instead of using the results from the fits to experimental data, which are affected by large uncertainties, one may have decided to use directly the purely theoretical results for $C_{7 \, (\perp,\|)}^{c\bar{c}}$ computed in Ref.\cite{Khodjamirian:2010vf} that are substantially smaller in absolute value. 
 In this sense the numbers shown for illustration in Tab.~\ref{tab:c7cc} can be considered as being rather conservative. 
 }.

\begin{table}
\small
\centering
\renewcommand{\arraystretch}{1.5}
\begin{tabular}{@{}c|c|c}
 & $C_{7 \, \perp}^{c\bar{c}}$ &  $C_{7 \, \|}^{c\bar{c}}$   \\
\hline 
$C_9^{\rm NP}=0$ & $-0.002\pm 0.102$ & $-0.133\pm 0.127$ \\
\hline
$C_9^{\rm NP}=-1.1$ & $+0.006\pm 0.084$ & $-0.147\pm 0.103$ 
\end{tabular}
\caption{Charm contribution entering $C_7$ as obtained from the fit for $n=1$ in the SM (Tab.~\ref{tab:A-SM}) and in presence of NP $C_9^{\rm NP}=-1.1$ (Tab.~\ref{tab:A-C9}).}
\label{tab:c7cc}
\end{table}

After discussing $C_7$, we can turn our attention to the other Wilson coefficients different from $C_9$, which are not  affected by long-distance charm contributions. 
The key observation is that some angular observables exhibit peculiar suppression mechanisms at low $q^2$ that protect them from contributions from $C_9$. One can identify three optimized observables of interest:
\begin{itemize}
\item $P_1$ and $P_3$ with a sensitivity to $C_7$ and $C_7^\prime$,
\item $P_2$ with a sensitivity to $C_7 C_{10}$ and $C_7^\prime C_{10}^\prime$.
\end{itemize}
These observables are protected from $C_9$ and its associated long-distance charm  (but obviously not from charm contributions to $C_7$) as they are built from the helicity amplitudes $A_{\perp,\|}^{L,R}$ that exhibit a photon pole contrary to the longitudinal amplitude $A_{0}^{L,R}$~\footnote{This also ensures that their computation in QCDF is infrared safe and thus under control even at very large recoil, as discussed in Ref.~\cite{Beneke:2001at}. Further alluring properties of these transverse asymmetries were discussed in Refs.~\cite{Kruger:2005ep, Becirevic:2011bp}.}. We will discuss now more precisely the mechanism at play for these observables.

\subsection{\boldmath$P_1$ and $P_3$ at very low $q^2$}

The observables $P_1$ (initially called $A_T^2$ in Ref.~\cite{Kruger:2005ep}) and $P_3$ (initially called $A_T^{\rm (Im)}$ in Ref.~\cite{Becirevic:2011bp}) are defined by
\begin{equation}
P_1=\frac{|A_\perp^L|^2+|A_\perp^R|^2 - |A_\|^L|^2- |A_\|^R|^2}{|A_\perp^L|^2+|A_\perp^R|^2 + |A_\|^L|^2+ |A_\|^R|^2} \quad {\rm and} \quad P_3=-\frac{{\rm Im}[A_\|^{L*} A_\perp^L + A_\perp^R A_\|^{R*}]}{|A_\perp^L|^2+|A_\perp^R|^2 + |A_\|^L|^2+ |A_\|^R|^2}\,.
\end{equation}
At very low $q^2$, $P_1$ and $P_3$ are sensitive  only to the electromagnetic coefficients  because $J_3$ has a double pole structure stemming from the photon pole. While $P_1$ is sensitive to  ${\rm Re}[C_7 C_7^\prime]$, $P_3$ depends on  ${\rm Im}[C_7 C_7^\prime]$. For simplicity and in agreement with the absence of significant $CP$-asymmetries in the current measurements, we will assume  that NP does not induce  new weak phases and only $C_9$ is complex, with an imaginary part due to SM effects only. We denote the different $C_9$ ($C_9^{c\bar{c}}$) contributions associated with each amplitude (following the notation of Eq.~(\ref{c9eff}))
\begin{eqnarray} C_{9 \, j}^{\,  R} \equiv &{\rm Re}\ C_{9 \, j}^{\rm eff }(q^2)&={C_{9 \, pert} ^{{\rm eff  \, SM} \, R}} + C_9^{\rm NP} + C_{9 \, j}^{c\bar{c}  \,\,  R}(q^2) \nonumber \\[1.0mm] C_{9 \, j}^{\,  I} \equiv
&{\rm Im}\ C_{9 \, j}^{\rm eff }(q^2)&={C_{9 \, pert}^{{\rm eff \, SM} \,  I}}  + C_{9 \, j}^{c\bar{c}    \,\, { I}}(q^2)\end{eqnarray}
with $j=\perp,\|,0$ and the superscript $R$ $(I)$ stands for Real (Imaginary) parts. In a similar way,  the Wilson coefficient $C_7$ can be written as
\begin{equation}  
C_{7\, \perp,\|}=C_7^{\rm eff \, SM}+ C_7^{\rm NP}+  C_{7\, \perp,\|}^{c\bar{c}}
\end{equation}  
where $ C_{7 \, \perp,\|}^{c\bar{c}}$ is an amplitude-dependent long-distance charm contribution associated to this coefficient and given in terms of helicity amplitudes in Eq.~(\ref{c7charm}).

Under this hypothesis of only real NP contributions, $P_3$ does not carry relevant information (see below). 
$P_1$ can be expanded in powers of $\hat{s}=s/m_B^2$ (with $\hat{m}_b=m_b/m_B$):
\begin{eqnarray} P_1&=&\frac{1}{N} \Bigg[ (2 C_7^\prime - C_{7\,\|} + C_{7\,\perp}) (C_{7 \, \perp}+C_{7 \, \|})/2 \\
&&\qquad\qquad +\left( C_{7 \,\perp} C_{9 \,\perp}^R - C_{7\,\|} C_{9 \,\|}^R + C_9^\prime (C_{7\,\perp} + C_{7\, \|})
  + C_7^\prime (C_{9\, \perp}^{\, R}+C_{9 \, \|}^{\,  R})  \right)  \frac{\hat{s}}{2 \hat{m_b}}+ \ldots\Bigg]\nonumber
\end{eqnarray}
where 
\begin{eqnarray}\label{nnn} N&=& \Bigg[ (C_{7 \, \perp}^2+ C_{7 \, \|}^2)/2+ C_7^{\prime \,2} + C_7^\prime( C_{7 \,\perp}-C_{7 \, \|}) \\
&&\qquad +\left( C_{7 \, \perp} C_{9 \, \perp}^R + C_{7 \,\|} C_{9\, \|}^R+ C_9^\prime (C_{7\, \perp}-C_{7\, \|})
+ C_7^\prime (C_{9 \, \perp}^{ \, R}-C_{9\, \|}^{\, R} + 2 C_9^\prime) \right) \frac{\hat{s}}{2 \hat{m_b}}+ \ldots \Bigg] \nonumber
\end{eqnarray}
The ellipsis denotes higher orders in  the expansion in $\hat{s}/(2 \hat{m_b})$. We have not combined the expansions of the numerator and denominator for simplicity of the discussion. As can be seen from this expansion, the contamination  from $C_9$ is suppressed at very-low ${\hat{s}}$ (for $s\leq 1$ GeV$^2$, $\hat{s}\leq 0.04$). Long-distance charm pollution from $C_7$ at very-low dilepton mass is present in both numerator and denominator, but it is expected to be small according to our discussion at the beginning of Sec.~\ref{sec:kinem}. The determination of $C_7$ from $P_1$ is unlikely to become competitive with the extraction from $b \to s \gamma$ decays.

On the contrary, in the absence of NP with imaginary contributions, $P_3$ becomes uninteresting (in the sense discussed in this section) since the leading term  is kinematically suppressed and  doubly contaminated by (the imaginary part of) $C_9$ and by charm inside $C_7$:
\begin{equation}
P_3 \propto \hat{s} \left[ C_{7 \, \perp} C_{9 \, \|}^I - C_{7 \, \|} C_{9 \, \perp}^I + C_7^\prime (C_{9\, \|}^{\, I}+C_{9\, \perp}^{\, I})\right] + \ldots
\end{equation}

\subsection{\boldmath$P_2$ at very low $q^2$}

The observable $P_2$ (originally called $A_T^{\rm (Re)}$ in Ref.~\cite{Becirevic:2011bp}) defined as 
\begin{equation} P_2= \frac{{\rm Re}[A_\|^L A_\perp^{L*} - A_\perp^R A_\|^{R*}]}{|A_\perp^L|^2+|A_\perp^R|^2 + |A_\|^L|^2+ |A_\|^R|^2}\end{equation}
involves all  Wilson coefficients $C_7^{(\prime)}$, $C_{9,10}^{(\prime)}$. At very low $q^2$, one would naively expect a behaviour similar to $P_{1,3}$, with a sensitivity to $C_7^{(\prime)}$ and a suppression of the semileptonic $C_{9,10}$ coefficients. Actually one finds that $P_2$ is independent of $C_9$ in this range but does exhibit a sensitivity to $C_{10}$.
Contrary to $P_{1,3}$, this sensitivity comes from a cancellation between left- and right-handed contributions in the numerator, which eliminates the double pole involving only electromagnetic operators and leaves the single pole as the dominant term. The same cancellation removes the sensitivity to the $C_9$ coefficient in the leading term. In the denominator the double pole survives and, as a consequence, the observable is  globally suppressed by  $\hat{s}$.
This can be seen analytically by expanding the observable in the large-recoil limit:
\begin{eqnarray} \label{p2eq}
P_2&=&\frac{\hat{s}}{4 \hat{m_b} N} \Bigg[  C_{10} (C_{7 \,\perp}+C_{7 \,\|})+  C_{10}^\prime (-2 C_7^\prime -C_{7 \,\perp}+C_{7 \,\|})    \\
&& \qquad\qquad +\left( C_{10} (C_{9 \,\perp}^{\, R}+C_{9\,\|}^{\, R}) + C_{10}^\prime (C_{9\,\|}^{\, R}-C_{9\, \perp}^{\, R}- 2 C_9^\prime) \right)
\frac{\hat{s}}{2 \hat{m_b} }\, \Bigg]+ \ldots  \nonumber
\end{eqnarray}
In the numerator, the contributions from $C_{9,(\perp,\parallel)}^{R} $ are suppressed by $\hat{s}$ with respect to the leading $C_7^{(\prime)} C_{10}^{(\prime)}$ contribution. In the denominator, given in Eq.(\ref{nnn}), the contributions from $C_{9,(\perp,\parallel)}^{R}$ and $C_9^\prime$ are always suppressed by $\hat{s}$. We have checked that this remarkable behaviour does not occur for other optimized observables:
for instance, we find a very similar situation in the numerator of $P_5^\prime$ (with a factor  $m_b m_B$) but its denominator exhibits no suppression of $C_9^{\rm eff}$ at small $\hat{s}$.

For small $\hat{s}$ (in particular the first bin [0.1,0.98] GeV${}^2$), $P_2$ is protected from contributions due to $C_9^{\rm eff}$ coming either from Standard Model, charm-loop, ad-hoc non-factorisable power corrections or New Physics. On the contrary, it is  sensitive to the product $C_7 C_{10}$ and the corresponding chirally flipped ones. Then from Eq.~(\ref{p2eq}) the leading term in $\hat{s}$ in the numerator of $P_2$ is of the form 
\begin{equation} \label{main} C_7^{\rm eff \, SM} C_{10}^{\rm SM} + C_7^{\rm NP} C_{10}^{\rm SM} + C_7^{\rm eff \, \rm SM} C_{10}^{\rm NP}   + \Delta C_7^{c\bar{c}}   +
 C_7^{\rm  NP} C_{10}^{\rm NP}- C_7^\prime C_{10}^\prime,
\end{equation}
where the first term is large and positive, the second and third term are numerically subleading, the last two terms are even more suppressed, and finally the term
\begin{equation}\label{deltac7} 
\Delta C_7^{c\bar{c}}=C_{10} ( C_{7 \, \perp}^{c\bar{c}}+ C_{7 \, \|}^{c\bar{c}})/2 + C_{10}^\prime ( C_{7 \, \|}^{c\bar{c}}- C_{7 \, \perp}^{c\bar{c}})/2
\end{equation}
 collects all long-distance charm contributions. Focusing first on the numerator of $P_2$, one can see that
 improving the agreement with the current LHCb data would require
\begin{equation} \label{npx} 
C_7^{\rm NP} C_{10}^{\rm SM} + C_7^{\rm eff \, SM} C_{10}^{\rm NP} + \Delta C_7^{c\bar{c}} \leq 0.
\end{equation}

Given that $|C_{10}^\prime| \ll |C_{10}|$ according to the global fit in Ref.~\cite{Descotes-Genon:2015uva}, one can safely neglect the right-handed currents in Eq.~(\ref{deltac7}). According to Tab.~\ref{tab:c7cc}, we see that  this  long-distance charm  term $\Delta C_7^{c\bar{c}}$ is positive in most of the 1 $\sigma$ range. Assuming no sizeable right-handed currents and taking into account both the numerator and the denominator one finds that a positive (negative) $\Delta C_7^{c\bar{c}}$ decreases (increases) the value of the first bin of the observable $P_2$ with respect to the SM  by a factor $(1-\Delta C_7^{c\bar{c}}/(C_7^{\rm eff \, SM} C_{10}))$. Using central values of Tab.~\ref{tab:c7cc}, the value of $P_2^{\rm SM}$ is reduced in the first bin to $0.87\, P_2^{\rm SM}$ ($0.83 \, P_2^{\rm SM}$) for $C_9^{\rm NP}=-1.1$ ($C_9^{\rm NP}=0$ respectively), when including these charm contributions (a much smaller effect is observed if the values of Ref.\cite{Khodjamirian:2010vf} for $C_{7 \, (\perp,\|)}^{c\bar{c}}$  are used instead).

\begin{figure} 
\includegraphics[width=7.7cm,height=8cm]{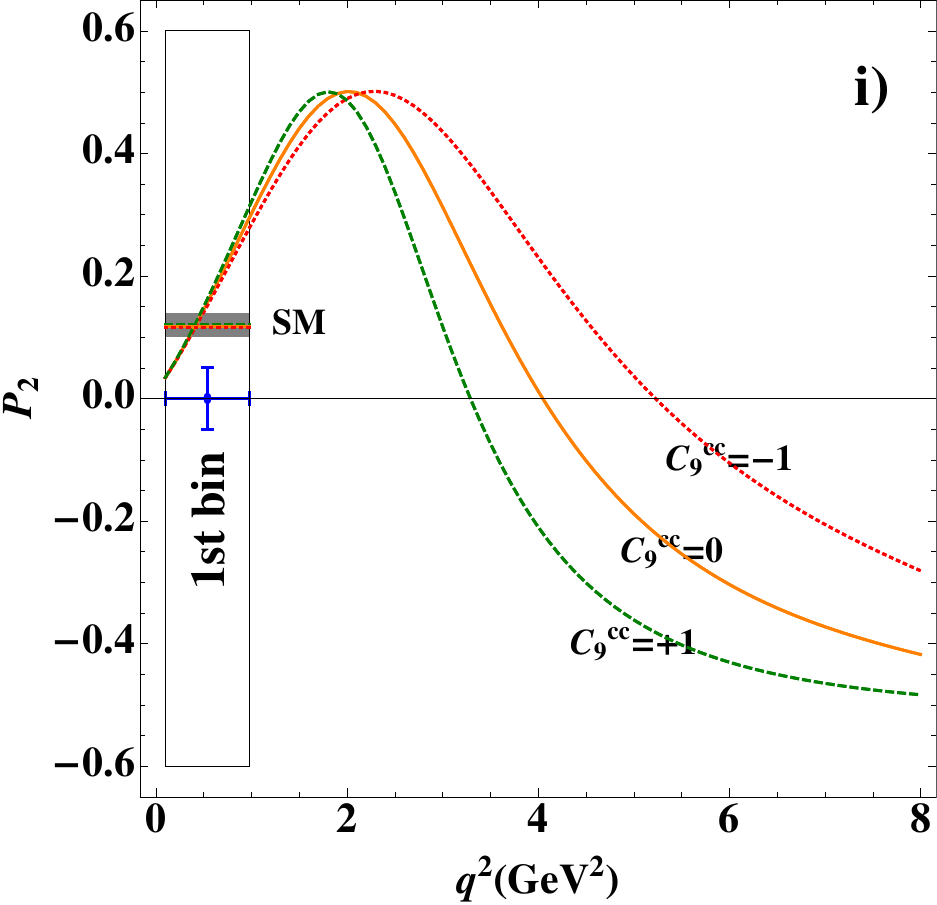}
\includegraphics[width=7.7cm,height=8cm]{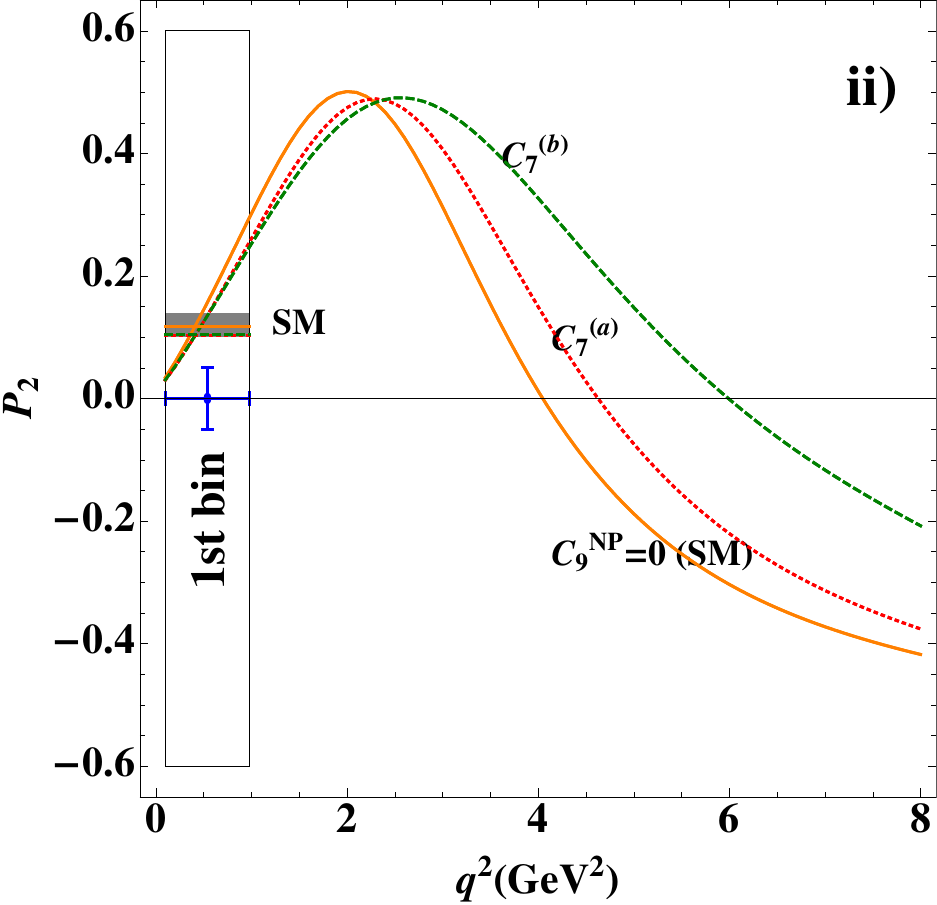}
\caption{i) Sensitivity of $P_2$ (central value only) to the variation of charm in $C_9$ taking $ C_9^{c\bar{c}}=C_{9 \, \perp}^{c\bar{c}}= C_{9 \, \|}^{c\bar{c}}=\pm 1$.  ii) $P_2$ sensitivity to the variation of charm in $C_7$: 
The solid orange line is the SM central value, the dotted red line (a) corresponds to the central value for $ C_{7 \, \perp}^{c\bar{c}}=-0.007$, $ C_{7 \, \|}^{c\bar{c}}=-0.093$, $C_9^{\rm NP}=0$, the green dashed line (b) corresponds to  the central value for $C_{7 \, \perp}^{c\bar{c}}=+0.005$, $ C_{7\, \|}^{c\bar{c}}=-0.083$,  $C_9^{\rm NP}=-1.1$. The blue cross indicates the LHCb measurement of the first bin~\cite{Aaij:2015oid} (strictly speaking for $\hat{P}_2$). The grey band corresponds to the SM prediction, whereas the coloured lines correspond to the central value of the binned prediction for each scenario. \label{fig:P2charm}}
\end{figure}

In order to illustrate the charm sensitivity of $P_2$, in particular in the region of the first bin, we consider the impact of a (universal) charm contribution entering $C_9$~\footnote{Eq.~(\ref{p2eq}) shows that $P_2$ at low $q^2$ is essentially sensitive to averages of $C_{i \, \perp}^{c\bar{c}}$ and $C_{i \, \|}^{c\bar{c}})$ with $i=7,9$, so that taking different contributions for each transversity amplitude would lead to similar results.}. 
This is illustrated in the left panel of Fig.~\ref{fig:P2charm}. The right panel shows that the sensitivity to the charm contribution to $C_7$ yields a larger but still limited effect.

\begin{figure}
\begin{center} 
\includegraphics[width=7.7cm,height=8cm]{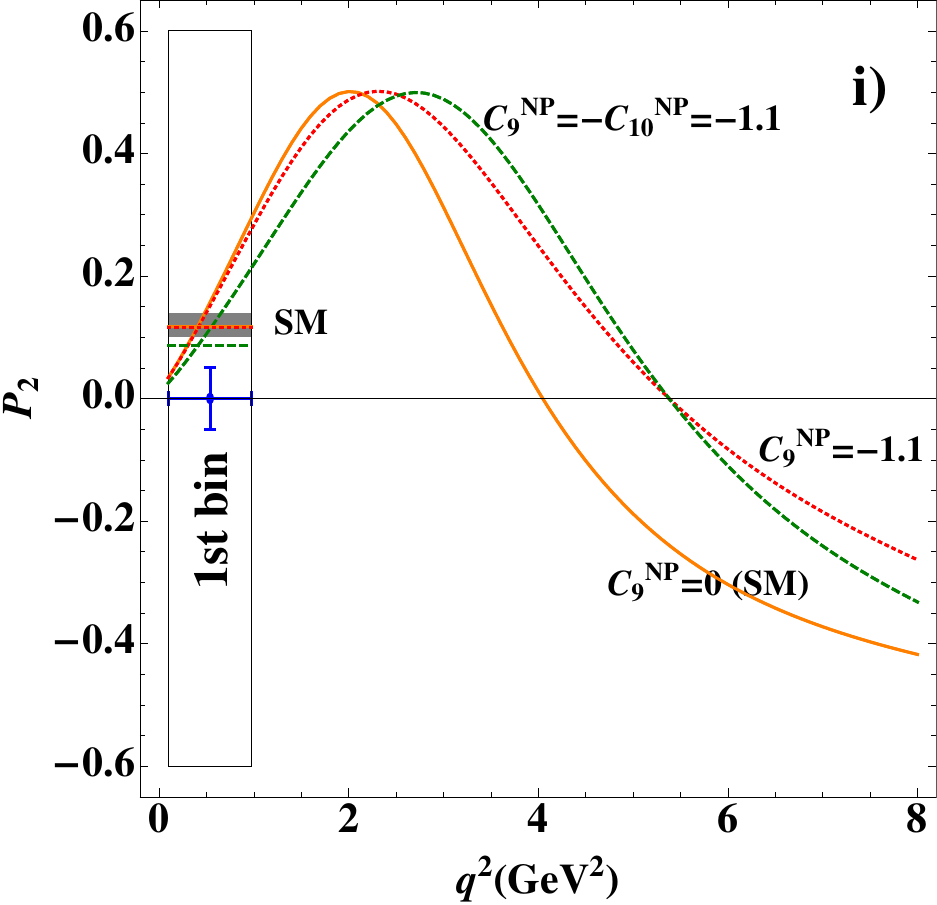}
\includegraphics[width=7.7cm,height=8cm]{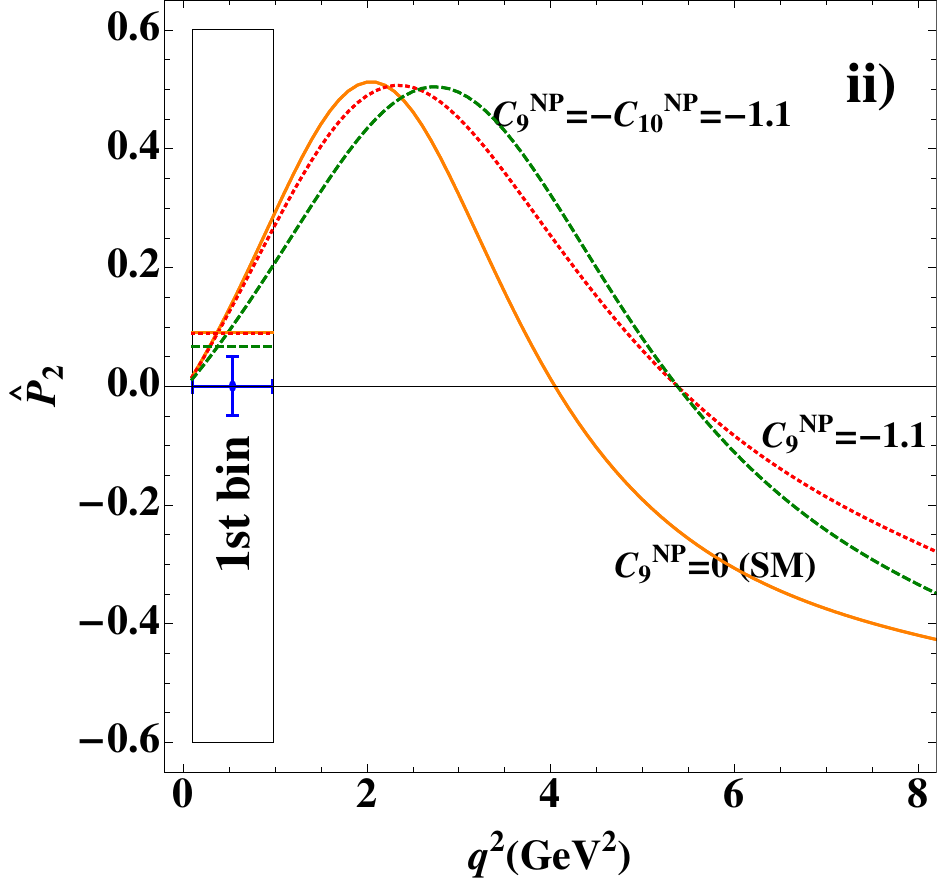}
\end{center}
\caption{i) $P_2$ sensitivity to NP. ii) $\hat{P_2}$ sensitivity to NP. Same conventions as Fig.~\ref{fig:P2charm}. 
\label{fig:P2NP}}
\end{figure}

The sensitivity of $P_2$ for different NP scenarios  is explored in Fig.~\ref{fig:P2NP}.
In agreement with Fig.~\ref{fig:P2charm}, the variations in $C_9$  (whether from charm or NP) are irrelevant for the first bin. On the other hand, a positive contribution in $C_{10}^{\rm NP}$ improves
the agreement between the prediction and data in the first bin. This contribution to $C_{10}$ also shifts the position of the maximum of $P_2$, but its zero. Let us remark that this shift of the maximum of $P_2$ 
(also produced by $C_9^{\rm NP}=-1.1$) would increase  the value of $P_2$ in the bin [2,4.3] GeV${}^2$ as observed in the LHCb 2013 data set (with a 2.9 $\sigma$ tension with respect to the SM).


A comment is in order concerning the comparison between data (blue crosses in Figs.~\ref{fig:P2charm} and \ref{fig:P2NP}) and theory in the first bin. Figs.~\ref{fig:P2charm} (left and right) and \ref{fig:P2NP} (left) show predictions for $P_2$. Due to the limited statistics, the LHCb analysis of the full $B\to K^*\ell\ell$ angular distribution is performed neglecting lepton mass effects, which corresponds to a change of the definition of the longitudinal polarisation $\hat{F}_L$ compared to the definition $F_L$ commonly used theoretically (see Sec.~2.3 in Ref.~\cite{Descotes-Genon:2015uva} for the definitions). Indeed, the measurement of $F_L$ is performed using $J_{1c}$, rather than $J_{2c}$ (used to define the optimized observables~\cite{Matias:2012xw,Descotes-Genon:2013vna}): both differ 
 by $m_\ell$-suppressed terms which are generally tiny, but noticeable at very low $q^2$.
An estimate of the impact of this approximation used by LHCb is shown in Ref~\cite{Descotes-Genon:2015uva}  and it was found to decrease the SM prediction of $P_2$ by around 23\% in the first bin compared to a computation based on $J_{2c}$.
This implies that LHCb does not measure  $P_2$ in this first bin but a modified observable, 
$\hat{P_2}$ \cite{Descotes-Genon:2015uva}. Numerically, in the case of interest analyzed here, we have found  that one can easily transform the theoretical values of $P_2$ into $\hat{P_2}$ using
 $ \av{\hat{P_2}}_{[0.1,0.98]}\simeq 0.77 \av{{P_2}}_{[0.1,0.98]} $ (as in the SM case). In Fig.~\ref{fig:P2NP} (right), we show the variation of $\hat{P_2}$ in several scenarios. Once again, a positive NP contribution in $C_{10}$ contribution improves the agreement between data and prediction.

\subsection{The implications of the Belle measurement of $Q_5$}

\begin{figure}[t]
\begin{center}
 \includegraphics[width=7cm]{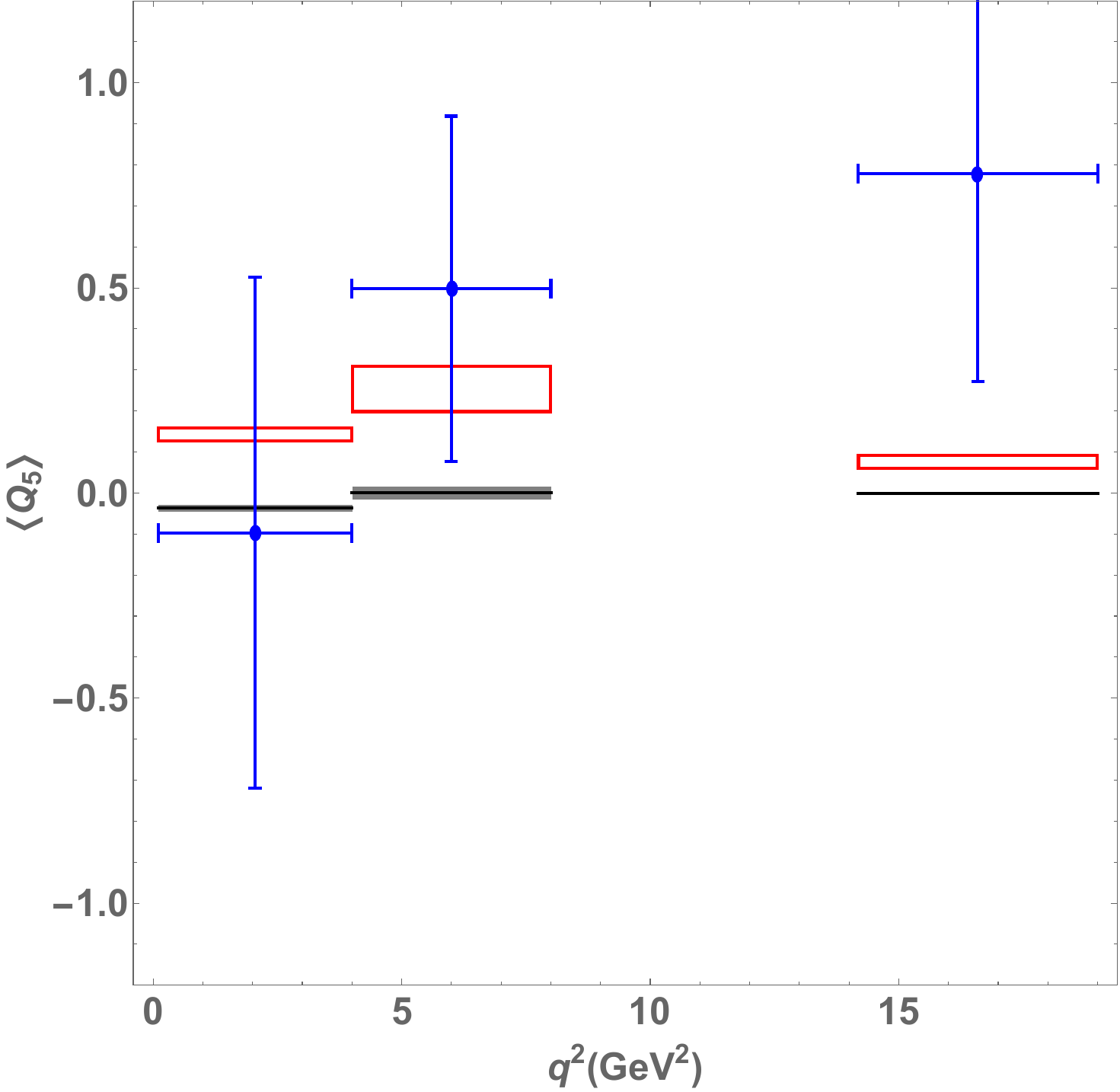}\quad\quad
  \includegraphics[width=7cm]{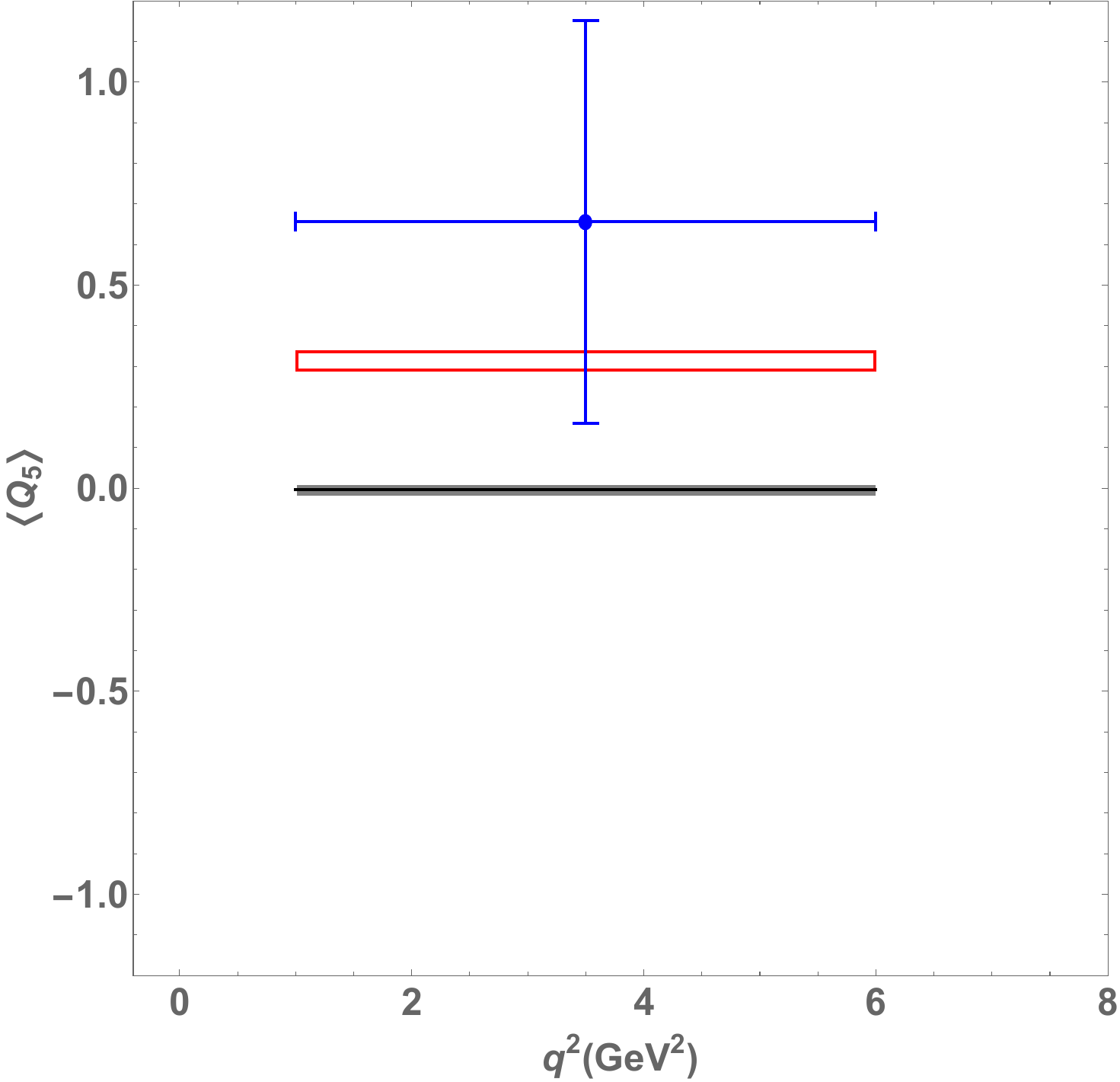}
    \end{center}
  \caption{Predictions for $Q_5$ in SM \cite{Capdevila:2016ivx} (black lines) and in presence of NP (red boxes) in $C_9^{\mu}=-1.1$ (with $C_9^e=0$) and data from Belle \cite{Wehle:2016yoi} (blue crosses).
   } \label{fig:P5}
\end{figure}

Our previous arguments show that neither factorisable power corrections nor charm loops 
are likely to account for the observed anomalies. In addition one can use a complementary and powerful independent tool to support these arguments, namely data.  
The recently proposed observable $Q_5=P_5^{\prime \mu} - P_5^{\prime e}$ \cite{Capdevila:2016ivx} hampers any possibility to use an SM alternative to explain the anomaly in $P_5^{\prime\mu}$. Independently of how large or of unknown origin or even wrong is the contribution added to the prediction of $P_5^{\prime \mu}$, {\it in the SM} the electronic $P_5^{\prime e}$ counterpart will receive the same contribution.
These SM contributions will automatically cancel in $Q_5$, up to contributions highly suppressed by $m_\ell^2$ and $q^2$ leading to extremely clean SM predictions (shown in Fig.~\ref{fig:P5}).

 Belle has been the first experiment to probe the observable $Q_5$~\cite{Wehle:2016yoi}: in the relevant bin [4,8] GeV$^2$, a good agreement with the LHCb measurement of $P_5^{\prime\mu}$~\cite{Aaij:2015oid} was observed,  with a 2.6 $\sigma$ deviation w.r.t. the SM prediction while only a 1.3 $\sigma$ deviation for the electronic observable $P_5^{\prime e}$ was found. This implies a 1.2 $\sigma$ deviation w.r.t. the SM for the corresponding observable $Q_5$ in the bin [4,8] GeV$^2$, which is reduced to 0.6 $\sigma$ in the presence of a NP contribution $C_9^{\mu}=-1.1$ (left-hand side of Fig.~\ref{fig:P5}). In the bin [1,6] GeV$^2$, one gets a discrepancy of 1.3 $\sigma$ in the SM, reduced to 0.7 $\sigma$ for a NP contribution $C_9^{\mu}=-1.1$ (right-hand side of Fig.~\ref{fig:P5}). 
 
The low statistical significance of this result prevents us from drawing any firm conclusion at this stage. It is however interesting to notice the similarities with the pattern observed in $R_K$. Both LHCb and Belle-II should have the capacity  to implement this important test  and to provide a robust complementary test to the arguments discussed in this paper.

\section{Conclusions}\label{sec:concl}

Over the last few years, a coherent pattern of deviations has emerged in $b\to s\mu^+\mu^-$ decays, from LHCb and Belle measurements. These deviations and their correlations can be analysed in the effective Hamiltonian approach, 
as done in several global analyses of $b\to s\mu^+\mu^-$ and $b\to s e^+e^-$ modes~\cite{Descotes-Genon:2015uva,Altmannshofer:2014rta,Altmannshofer:2015sma,Hurth:2016fbr}. 
The outcome is intriguing: a shift in the Wilson coefficient $C_9^\mu$ by about -25\% of its SM value  is sufficient to
achieve a significant improvement (by more than 4 $\sigma$) in the description of the data (contributions to other coefficients like $C_{10}^\mu$ and $C_{9'}^\mu$ are also allowed). There have been several controversies concerning the assessment of theoretical uncertainties in the predictions of $B \to K^*\mu^+\mu^-$ observables: some concerning the factorisable QCD corrections arising in the description of form factors, whereas other dealt with non-factorisable corrections only present at the level of the amplitudes and related to long-distance charm-loop contributions. Even though these effects could not explain the 2.6 $\sigma$ anomaly in the ratio $R_K$~\cite{Aaij:2014ora}, which goes in the same direction as the (statistically not yet relevant) trend observed for the difference between $P_5'$ for electrons and muons~\cite{Wehle:2016yoi}, it is interesting to assess these claims concerning $B \to K^*\mu^+\mu^-$ observables.

The first discussion deals with factorisable corrections. In the limit $m_b \to \infty$, the seven $B\to K^*$ form factors can be reduced to two soft form factors $\xi_{||}$ and $\xi_\perp$, but these relations get corrected not only by computable perturbative corrections from hard gluons, but also by power corrections of $\mathcal{O}(\Lambda/m_B)$ (and higher). These power corrections must be modeled on the basis of dimensional estimates. Moreover, a choice must be made to determine $\xi_{||}$ and $\xi_{\perp}$ from 
non-perturbative input (typically obtained from light-cone sum rules). This is done by identifying these soft form factors with (combinations of) full form factors,
and thus setting the corresponding power corrections to zero. There are several possible choices (``schemes'') for this identification, and we assessed the role played by the scheme prescription for the accuracy of the SM predictions for $B\to K^*\mu^+\mu^-$ observables.

We showed that, in the absence of further information on the correlations among form factors, the choice of scheme has an impact on the theoretical uncertainties for predictions. Uncertainties for observables can easily be overestimated by choosing an inappropriate choice of scheme, for instance if soft form factors are identified with full form factors  playing little to no role in the computation of these observables. We demonstrated the origin of this scheme dependence in a pedagogical way and derived analytic formulae for the contribution from power corrections to the most important optimized observables $P_5^\prime$, $P_2$ and $P_1$. We further showed that a fit of power corrections for the scheme used in Ref.~\cite{Descotes-Genon:2015uva} to the form factor input from BSZ~\cite{Straub:2015ica} yields uncertainties associated to power corrections in agreement with
the generic 10\% dimensional estimate as expected. We compared predictions for $P_5^\prime$ with uncorrelated power corrections and soft form factors to those where correlations are assessed from BSZ form factors, 
and we established that the main source of correlations among form factors comes from the symmetry relationships in the $m_b\to\infty$ limit, 
whereas the correlations among power corrections are subleading effects. 
Our findings disprove claims of significantly larger uncertainties from factorisable power corrections
made in Refs.~\cite{Jager:2012uw,Jager:2014rwa}. 

Concerning non-factorisable QCD corrections related to long-distance charm loops, the problem to disentangle NP from a non-perturbative QCD effect is more complicated, although a handle is provided
by the expected non-trivial dependence on the squared dilepton-mass $q^2$ of the charm loop (and on the initial and final hadrons). The Wilson coefficient $C_9$ can be written in the particular case of $B \to K^*\mu^+\mu^-$ as
$C_{9 \, i}^{{\rm eff}}(q^2)={C_9^{\rm eff}}_{\rm SM pert.}+C_9^{\rm NP}+C_{9 \, i}^{{c\bar c} }(q^2)$, where $i$ labels the transversity of the lepton pair.
The perturbative SM and NP contributions are accompanied by a long-distance charm loop contribution $C_{9 \, i}^{c\bar c}(q^2)$.  In our analysis in Ref.~\cite{Descotes-Genon:2015uva} 
we included the partial LCSR computation from Ref.~\cite{Khodjamirian:2010vf} as an estimate of the order of magnitude of the functions $C_{9 \, i}^{c\bar c}(q^2)$. 
Recently, in Ref.~\cite{Ciuchini:2015qxb} several fits of the $C_{9 \, i}^{c\bar c}(q^2)$ to $B\to K^*\mu^+\mu^-$ measurements were performed and it was claimed that the data favoured a $q^2$-dependent 
contribution rather than a universal shift in $C_9$.
We re-analysed these claims and stressed that the $q^2$-dependence observed in some of the fits in Ref.~\cite{Ciuchini:2015qxb} was actually due to imposing a pure SM constraint from Ref.~\cite{Khodjamirian:2010vf}  at very large recoil, skewing the fit and generating an apparent $q^2$-dependence to get a better agreement with data at higher $q^2$. Moreover, we pointed out a mismatch 
in the identification of Ref.~\cite{Ciuchini:2015qxb} to the results of Ref.~\cite{Khodjamirian:2010vf}: the real parametrisation used in Ref.~\cite{Khodjamirian:2010vf} is matched to the modulus of the complex parametrisation adopted in Ref.~\cite{Ciuchini:2015qxb}.

We further stress that a potential $q^2$-dependence cannot be inferred from considering only the deviation of a single quantity among the large number of parameters entering the fits (as done in Ref.~\cite{Ciuchini:2015qxb}). 
The relevant issue consists in the improvement of the quality of the fit when going from the hypothesis of a constant $C_9$ (NP-like contribution)  to the hypothesis of a $q^2$-dependent $C_{9 \, i}^{{c\bar c}}$ (hadronic contribution).
Using the polynomial parametrisation of Ref.~\cite{Ciuchini:2015qxb} and the framework of Ref.~\cite{Descotes-Genon:2015uva}, we have performed the corresponding analysis using a frequentist statistical approach. We considered only $B\to K^*\mu^+\mu^-$ data, removed long-distance contributions estimated from Ref.~\cite{Khodjamirian:2010vf} and introduced a polynomial parametrisation
 describing charm-loop contributions with parameters to be fitted. We assumed either the SM value for the Wilson coefficients or we took $C_9^{\mu,{\rm NP}}=-1.1$, we used different form factors and approaches, and we even considered a fit including all available data on other $b\to s\mu^+\mu^-$ and $b\to s e^+e^-$ channels. 
In none of the scenarios there is a motivation to go beyond the linear order in the polynomial parametrisation (corresponding to a $q^2$-dependence closely equivalent to a constant contribution to $C_9$): even if in some cases one may get fits with quadratic terms different from zero, the improvement compared to the linear case is completely marginal~\footnote{Another group~\cite{nazilla} also reached similar conclusions following a different approach for their fits.}. These findings show that there is currently no indication for a non-trivial $q^2$-dependence for the $C_9$ contribution~\footnote{Recently, the authors of Ref.~\cite{Ciuchini:2015qxb} updated their analysis in Ref.~\cite{Ciuchini:2016weo}, stating that, in the case of a general fit without constraint, no conclusion on the presence of polynomial terms purely associated
with hadronic effects could be drawn.}, disfavouring 
an explanation of the $B\to K^*\mu^+\mu^-$ anomalies via non-factorisable QCD effects corresponding  to a charm-loop contribution with a pole at $q^2=m_{J/\psi}^2$.

Although we did not find an indication for underestimated hadronic uncertainties affecting the extraction of $C_9$ from global fits, we would like to stress that it is
important to assess also potential NP contributions to other Wilson coefficients, whose interpretations in terms of short-distance physics are not affected by  hadronic uncertainties. Indeed, the high sensitivity of a large set of observables to the Wilson coefficient $C_9$ pointing to a large tension with its SM value may have hidden contributions from the remaining semi-leptonic Wilson coefficients. Even if a global fit may constrain all Wilson coefficients simultaneously, some observables in specific regions may 
prove better adapted to track specific coefficients different from $C_9$ and potentially very interesting in terms of NP. In particular, we have discussed how $P_{2}$ for $B\to K^*\mu^+\mu^-$ at very low $q^2$  could provide further information on the Wilson coefficient $C_{10}$. 
A deviation from SM expectations for this observable can only be explained by NP in $C_{10}$, which cannot be mimicked by SM hadronic effects: charm-loop contributions to $C_7$ are constrained to be small from the comparison of inclusive and exclusive $b\to s\gamma$ decays, whereas $C_9$ contributions are suppressed for this observable in this kinematic range. Interestingly, a positive NP contribution to $C_{10}$ could improve the agreement between data and theory in the very low $q^2$ region. This approach complements  the one presented in Ref.~\cite{Capdevila:2016ivx}, which dealt with the case where lepton-flavour universality is violated (as suggested by the observables $R_K$, $R(D)$, $R(D^*)$): two observables $B_5$ and $B_{6s}$ provide then clean information on $(C_{10}^{\mu}-C_{10}^{e})/C_{10}^e$ (with no pollution from $C_7$).
 
We conclude with the obvious remark that the observation of deviations in optimized and lepton-flavour-violating observables like $Q_i=P_i^{\mu}-P_i^{e}$ would be an unambiguous signal of New Physics, rendering the discussion 
on hadronic explanations in Refs.~\cite{Jager:2012uw} and \cite{Ciuchini:2015qxb,Ciuchini:2016weo} irrelevant. A first step in this direction, albeit with a still limited statistical significance, 
is provided by the very recent results of the Belle experiment~\cite{Wehle:2016yoi}, which suggest that $P_5'$ would agree with the SM for electrons but disagree for muons, in the same direction as global fit results~\cite{Descotes-Genon:2015uva,Altmannshofer:2014rta,Altmannshofer:2015sma,Hurth:2016fbr}. Such exciting results call for more measurements from both LHCb and Belle-II collaborations in order to exploit the full potential of $b\to s\ell^+\ell^-$ transitions in the search for New Physics.

\section*{Acknowledgements}

We thank M.~Beneke, A.~Khodjamirian and J.Virto for fruitful discussions concerning the topics covered in this article.
This work received financial support from the grant FPA2014-61478-EXP [JM, SDG, BC, LH]; from the grants FPA2013-46570-C2-1-P, FPA2014-55613-P and 2014-SGR-104, 2014-SGR-1450 and from the Spanish MINECO under the project MDM-2014-0369 of ICCUB (Unidad de Excelencia Maria de Maeztu) [LH] and Centro de Excelencia Severo Ochoa SEV-2012-0234 [BC]; from the EU Horizon 2020 program from the grants No 690575, No 674896 and No. 692194 [SDG].

\appendix

\section{Parametric and soft form factor errors for $B\to K^*\ell\ell$ predictions}\label{App:A}

In this article, we have focused mainly on two sources of uncertainties: (factorisable) power corrections and (non-factorisable) charm-loop contributions.
For completeness, we discuss here the size of other error sources computed in different articles, CJ12~\cite{Jager:2012uw}, DHMV14~\cite{Descotes-Genon:2014uoa}, BSZ15~\cite{Straub:2015ica} and BBD14~\cite{Beaujean:2013soa}, 
considering an observable predicted in all papers:  $\av{P_5^\prime}_{[1,6]}$.
Given that parametric and form factor errors are not separated in some of the papers we will add them in quadrature for this comparison. The result is shown in Tab.~\ref{tab:P5p16} where  also factorisable power correction errors in this bin  are given.

\begin{table}
\small
\centering
\renewcommand{\arraystretch}{1.5}
\begin{tabular}{@{}c|c|c|c|cc@{}}
$\av{P_5^\prime}_{[1,6]}$ & DMHV14 & BSZ15  &  BBD14 & CJ12   \\
\hline
{$\Delta_{\rm par+FF}$}  & $\pm 0.066$ & $\pm 0.035$ &\multirow{2}{5em}{$<\pm 0.09^*$} & 
$\pm 0.12$ \\
\cline{1-3}  \cline{5-5} 
{$\Delta_{\rm factorisable\, p.c.}$}  & $\pm 0.093$ & - &  & $\pm 0.24$ \\
\end{tabular}
\caption{Uncertainties on the SM prediction for $P_5^\prime$ in the long bin $[1,6]$\,GeV$^2$: the first row gives  the parametric and form factor uncertainty added in quadrature, the second row provides the error from  factorisable power corrections. In the BBD14 case~\cite{Beaujean:2013soa}, only the total error size is given for nominal  priors, suggesting that this number should be taken as an upper bound of the subset of errors discussed here.}
\label{tab:P5p16}
\end{table} 

The parametric and soft form factor errors in DHMV14~\cite{Descotes-Genon:2014uoa} were computed by performing a random flat scan of all relevant parameters (masses, decay constants, renormalization scale ...)  within their uncertainty, keeping all other parameters (form factors, power corrections) to their central values. Then the observables are computed at each point of the scan and  their error bars were obtained in DHMV14~\cite{Descotes-Genon:2014uoa} computing the difference between the extreme values obtained for the observables in the scan with respect to the central value of the observable. 
The corresponding scan of parameters in BSZ15~\cite{Straub:2015ica} yields smaller errors than the ones in DHMV14 due to the much smaller uncertainties of the form factor inputs and the Gaussian treatment of all errors in BSZ15. Let us also remark that the  {\it total error} in BBD14~\cite{Beaujean:2013soa} in the nominal-prior evaluation is  in the same ballpark as the one in DHMV14. 

On the other hand, it is interesting to notice that the parametric uncertainty (including form factors) in CJ12~\cite{Jager:2012uw} is 2 to 3 times larger than the one in DHMV14~\cite{Descotes-Genon:2014uoa} and BSZ15~\cite{Straub:2015ica}, respectively. 
This issue is independent of and adds to the  inflation of errors associated to factorisable power corrections by a factor of 2 
due to the choice of scheme, as is discussed in Sec.~\ref{sec:BSZfit} and can be seen in the second row of Tab.~\ref{tab:P5p16}.
Let us also mention that in a subsequent article (CJ14, Ref.~\cite{Jager:2014rwa}) by the same authors,
the total error for the same bin  increased by 40\%  with respect to to the previous prediction in CJ12. In a
later article from the Belle collaboration~\cite{Abdesselam:2016llu}, the prediction for the same quantity, provided by one of the authors of CJ12 and CJ14,
got an uncertainty reduced by 60\% compared to CJ14 (see Table VI of Ref.~\cite{Abdesselam:2016llu}).
Unfortunately the absence of a precise error budget in Refs.~\cite{Jager:2014rwa,Abdesselam:2016llu} prevents us from exploiting 
the corresponding results for our comparison. Moreover, we are not in a position to explain the origin of the 40\% increase and subsequent 60\% decrease in these two articles, which is unfortunately not commented on in either case. 

One might suspect that the origin of this large difference between the 
error attached to  parametric and soft form-factor uncertainties in DHMV14, BSZ15 and BBD14 on one side and CJ12 and CJ14 on the other  could be the error attached to the soft form factor. However, the uncertainty for $\xi_\perp(0)=0.31 \pm 0.04$ in CJ14, estimated by considering {\it only the central values} of different form factor determinations, is even significantly smaller (by a factor around 4) than the one for $\xi_\perp(0)=0.31^{+0.20}_{-0.10}$ in DHMV14
from the calculation in Ref.~\cite{Khodjamirian:2010vf}.

In summary, in addition to the inflated power correction error related to an inappropriate choice of scheme, discussed in Sec.~\ref{sec:BSZfit}, we conclude that the analysis of the parametric errors in CJ12 is at odds with the results of three different groups (DHMV14, BSZ15, BBD14). 

\section{Predictions for $R_{K^*}$ in various scenarios}\label{App:B}

As discussed in the introduction and in section 5.3, it is of utmost importance to have observables able to test lepton-flavour universality. Among this type of observables, $R_K$~\cite{Aaij:2014ora} and the recently measured $Q_5$~\cite{Wehle:2016yoi} are already providing very interesting information. Following the structure of $R_K$ one can construct observables with similar capacities for other channels. Because of the anomalies observed in the $B\to K^*\mu\mu$ mode~\cite{Aaij:2015oid,Aaij:2013aln,Aaij:2015esa}, the observable $R_{K^*}={\cal B}_{B \to K^* \mu^+\mu^-}$ $/{\cal B}_{B \to K^* e^+e^-}$~\cite{Hiller:2003js} becomes a natural candidate to analyse. In this appendix, we provide our predictions for $R_{K^*}$ in three different bins both in the context of the SM and considering several NP scenarios suggested by global fits~\cite{Descotes-Genon:2015uva}.

\newpage

\begin{table}[h]
\begin{center}

\begin{tabular}{@{}cccc@{}}
\toprule[1.6pt] 
\multicolumn{4}{c}{$R_{K^*}$ Predictions}\\
\midrule 
 & $[0.045,1.1]$ & $[1.1,6.]$ & $[15.,19.]$\\
\midrule \medskip
Standard Model & $0.922 \pm 0.022$ & $1.000 \pm 0.006$ & $0.998 \pm 0.001$ \\ \medskip
$C_{9\mu}^\text{NP}=-1.1$ & $0.904 \pm 0.053$ & $0.868 \pm 0.082$ & $0.788 \pm 0.004$ \\ \medskip
$C_{9\mu}^\text{NP}=-C_{10\mu}^\text{NP}=-0.65$ & $0.869 \pm 0.065$ & $0.738 \pm 0.027$ & $0.701 \pm 0.006$ \\ \medskip
$C_{9\mu}^\text{NP}=-C_{9^\prime\mu}^\text{NP}=-1.07$ & $0.872 \pm 0.094$ & $0.783 \pm 0.138$ & $0.698 \pm 0.015$ \\
\begin{minipage}{4.5cm} \centering $C_{9\mu}^\text{NP}=-C_{9^\prime\mu}^\text{NP}=-1.18$ \\ $C_{10\mu}^\text{NP}=C_{10^\prime\mu}^\text{NP}=0.38$ \end{minipage} & $0.871 \pm 0.095$ & $0.745 \pm 0.120$ & $0.658 \pm 0.014$ \\ 
\bottomrule[1.6pt] 
\end{tabular}

\end{center}
\caption{Predictions for $R_{K^*}={\cal B}_{B \to K^* \mu^+\mu^-}$ $/{\cal B}_{B \to K^* e^+e^-}$ in the SM and various NP scenarios.}
\end{table}

\end{document}